\documentclass[a4paper,11pt]{article}
\pdfoutput=1 

\usepackage{jheppub} 
\usepackage{graphicx} 
\usepackage{upgreek}
\usepackage{amsmath}
\usepackage{amssymb}
\usepackage{multirow}
\usepackage{appendix}

\newcommand{\nn}{\nonumber \\}
\newcommand{\es}{& = &}

\newcommand{\ps}{& + &}
\newcommand{\ms}{& - &}

\newcommand{\3}{\mbox{${\bf \underline{3}}$}}
\newcommand{\s}{\mbox{${\bf \underline{1}}$}}
\newcommand{\spr}{\mbox{${\bf \underline{1}'}$}}
\newcommand{\sppr}{\mbox{${\bf {\underline{1}''}}$}}
\newcommand{\be}{\begin{eqnarray}}
\newcommand{\ee}{\end{eqnarray}}
\newcommand{\Pdag}{\phi^\dagger_i}
\newcommand{\pdag}{\phi^\dagger_0}
\newcommand{\PdagP}{\phi^\dagger_i \phi_i}
\newcommand{\psq}[1]{\phi_{#1}^2}
\newcommand{\pdagp}[2]{\phi_{#1}^\dagger\phi_{#2}}

\begin{document} 

\title{Features of Charged Lepton Flavor Violation in an $A_4$ Symmetric Neutrino Mass Model}
\author[1]{Pravesh Chndra Awasthi,} 
 \author[1]{Jai More,}
 \author[1]{Akhila Kumar Pradhan,}
 \author[1]{Kumar Rao,}
 \author[1]{Purushottam Sahu}
\author[1,2]{and S. Uma Sankar}
\affiliation[1]{Department of Physics, Indian Institute of Technology Bombay, Powai, Mumbai 400076, India}
\affiliation[2]{Raja Ramanna Professor, Institute of Mathematical Sciences, Taramani, Chennai 600115, India.}
\emailAdd{214120016@iitb.ac.in}
\emailAdd{more.physics@gmail.com}
\emailAdd{akhilpradhan@iitb.ac.in}
\emailAdd{kumar.rao@phy.iitb.ac.in}
\emailAdd{purushottam.sahu@iitb.ac.in}
\emailAdd{uma@phy.iitb.ac.in}

\abstract{Neutrino flavour oscillations imply that there must be charged lepton flavour violation (CLFV) also. Different neutrino mass models predict different patterns of CLFV decays. Analysis of this pattern in standard seesaw mechanisms shows that the amplitude for meson CLFV decay is correlated to the amplitude of the corresponding radiative CLFV decay. This correlation leads to the prediction that the branching ratios of meson CLFV decay will always be smaller than those of the radiative CLFV decays. In this work, we analyse an interesting neutrino mass model, based on $A_4$ symmetry, in which the symmetry and the symmetry-breaking pattern lead to the neutrino mixing matrix to be of tri-bimaximal (TBM) form. In this model, we find that the meson CLFV decay amplitudes are not correlated to the corresponding radiative CLFV decay amplitudes, unlike in the case of seesaw models. The branching ratios of radiative CLFV decays are predicted to be negligibly small in this model, but those of the meson CLFV decays can be large enough to be observable in the near future.
}

\maketitle

\section{Introduction}
The Standard Model (SM), in its minimal form, predicts neutrinos to be massless, which leads to exact lepton flavour conservation. Neutrino flavour oscillations have provided convincing evidence that exact flavour conservation does not hold. Theoretically, flavour oscillations are explained assuming that neutrinos have tiny masses and their flavour eigenstates mix. To account for neutrino masses and mixings, we must have physics beyond the Standard Model (BSM). It is also expected that the mass generation mechanism of the neutrinos is quite different from that of the charged fermions because the neutrino masses are at least six or seven orders of magnitude smaller. Since the neutrino of a given flavour forms an $SU(2)_L$ doublet with the charged lepton of the same flavour,  flavour violation in the neutrino sector necessarily leads to flavour violation in the charged lepton sector. Hence, charged lepton flavour violating (CLFV) decays, such as $\mu \to e \gamma$, $\mu \to 3 \, e$, and $K_L \to \mu \,e$, must have non-zero branching ratios. So far, no experimental evidence is found for CLFV decays, but a number of experiments are actively searching for them \cite{MEGII:2023ltw, COMET:2018auw, Mu3e:2020gyw}. If observed, it will be a very important signal for physics beyond the Standard Model.

The rates for various CLFV decays as well as the interrelations between them depend very strongly on the details of the neutrino mass model \cite{King:2003jb}. A detailed study of CLFV processes can thus provide a critical insight into the underlying structure of the neutrino mass generation mechanism. Most of the neutrino mass models utilise the seesaw mechanism to obtain tiny neutrino masses, without excessive fine-tuning. There are three popular implementations of the seesaw mechanism, which are labeled type-I, type-II, and type-III seesaw.
In type-I and type-III seesaw models, the loop induced CLFV amplitudes are proportional to the mixing between the light neutrinos and the heavy neutral fermions. This mixing is inversely proportional to the heavy fermion mass and is expected to be small, which causes all the CLFV decay branching ratios to be small. It must be noted that, in these scenarios, the branching ratios of meson CLFV decays, such as $K_L \to \mu\,e$, should be smaller than those of the corresponding radiative CLFV decays, such as $\mu \to e\, \gamma$, because of the additional suppression due to quark mixing factors from the Cabibbo-Kobayashi-Maskawa (CKM) matrix \cite{Awasthi:2024nvi}. 

At present, neither radiative CLFV decays nor meson CLFV decays are observed. A priori, it is possible for a meson CLFV decay to have a larger branching ratio compared to the corresponding radiative CLFV decay. If such were the case, neutrino mass generation must be more complex than the simple seesaw mechanism. In such models, the meson CLFV decays will not be correlated to corresponding radiative CLFV decays. In this work, we study one such model \cite{He:2006dk}.

There are a number of neutrino mass models proposed in the literature \cite{King:2003jb}. Most of them are based on various discrete flavour symmetries, which help in generating the observed patterns in neutrino masses and mixings. A number of popular models are based on the non-Abelian group $A_4$. The model proposed in Ref~\cite{He:2006dk} is particularly attractive because the Pontecorvo-Maki-Nakagawa-Sakata (PMNS) mixing matrix for the light neutrinos takes the TBM form due to the pattern of symmetry and symmetry breaking. However, TBM form of the PMNS matrix predicts $\theta_{13}$ to be zero, whereas the experiments measure it to be $\theta_{13} \approx 8.5^\circ$ \cite{DayaBay:2012fng, RENO:2012mkc, Esteban:2024eli}. It is possible to generate a viable value of $\theta_{13}$, in the above model by introducing a small perturbation to the Majorana mass matrix of heavy right-handed neutrinos \cite{Dev:2015dha}. It is interesting to note that this perturbation also leads to maximal CP violation.

This model contains many scalar multiplets, each with a different set of quantum numbers, which in general lead to tree-level flavour-changing neutral couplings (FCNC) of heavy neutral scalar bosons. We show that the $A_4$ symmetry of this model constrains the FCNCs such that both neutral meson mixing and radiative CLFV decays, due to these couplings, are forbidden. However, the allowed FCNCs can lead to some very distinctive meson CLFV decays. Thus, we have one good example of a neutrino mass model in which the meson CLFV decay amplitudes are not correlated to the corresponding radiative CLFV decay amplitudes. 

In ref.~\cite{Korrapati:2020rao}, we did a preliminary study of the purely leptonic CLFV decays of mesons in this model, using a very simplified version of the scalar potential. This simplification led to a set of massive complex neutral scalars, two of which were degenerate and had only flavour violating couplings to charged fermions. The degeneracy, combined with the structure of the flavour violating couplings, allowed a non-zero tree-level amplitude for $K^0 \to \mu^+ e^-$ but forced the tree-level amplitude for $K^0 \to \mu^- e^+$ to be zero. Similar charge selection rules held for CLFV decays of other neutral pseudo-scalar mesons.
 
In this work, we consider the full scalar potential, with the assumption that it is CP-conserving. When all the terms in the scalar potential are taken into account, we cannot consider the neutral scalar mass eigenstates as complex scalars. We construct the mass matrices for both the real neutral scalars and pseudo-scalars and identify the mass eigenstates.
We find that the couplings of fermions to one of the real scalars are the same as those to the SM Higgs. Of the remaining three, one has only flavour-conserving couplings to charged fermions, while the other two have degenerate masses and have only flavour-changing couplings to charged fermions. Among the pseudo-scalars, one behaves like a Goldstone boson, whereas two of the remaining three have degenerate masses and have the same pattern of flavour-violating couplings as the degenerate real scalar fields. We find that the structure of the flavour-violating couplings of neutral scalars (both scalar and pseudo-scalars) obtained here is the same as the one for the complex neutral scalars, obtained in ref. \cite{Korrapati:2020rao} with the very simplified form of the potential. Because of the degeneracy of masses of the neutral scalars and the structure of their flavour-violating couplings to charged fermions are the same in both the simplified analysis as well as the full analysis, the pattern of CLFV decays of neutral pseudo-scalar mesons is also the same, which is a consequence of the $A_4$ symmetry of the potential in both cases. This naturally leads to the same constraints on the masses of the neutral scalars with flavour-violating couplings in the two cases. Unlike in ref. \cite{Korrapati:2020rao}, which considered only purely leptonic CLFV decays, 
here we include the semileptonic CLFV decays as well.

The paper is organised in the following way: In 
sec.~\ref{description}, we give a short description of the model of ref.~\cite{He:2006dk} and the detailed expression for its Yukawa Lagrangian, fermion masses and mixings. The light neutrino masses in this model arise through type-I seesaw mechanism. We work out the values of the Dirac and Majorana masses of the neutrinos which satisfy the constraints from neutrino oscillations and show that the light-heavy neutrino mixing is of the $\mathcal{O}(10^{-7})$, which leads to negligibly small loop-induced CLFV amplitudes \cite{Awasthi:2024nvi}. In sec.~\ref{ScalarPot}, we write down the full scalar potential and diagonalise  the matrix of the neutral real scalars to derive the orthogonal matrix which connects the mass eigenstates to $A_4$ eigenstates. A similar procedure is implemented for the neutral pseudo-scalars also. We also rewrite the Yukawa Lagrangian of the model in the mass eigenbasis of the fermions and of the scalars. In sec.~\ref{CLFVdecays}, we study all the possible CLFV decays of mesons and other possible CLFV processes. We present our conclusions in sec.~\ref{conclusion}.

\section{About the model}\label{description}
\subsection{A brief description of the model}
The model we consider was proposed in Ref~\cite{He:2006dk}.
It is based on the Standard Model (SM) gauge group but also has an additional $A_4$ discrete symmetry. The group $A_4$, the alternating group of order twelve, consists of all even permutations of four elements. It features three one-dimensional representations, $\s, \spr$ and $\sppr$, along with a three-dimensional representation, $\3$. A very attractive feature of this model is that it generates a 
TBM form for the Pontecorvo-Maki-Nakagawa-Sakata
(PMNS) matrix purely from symmetry and symmetry breaking.

The charged fermion content of this $A_4$ model \cite{He:2006dk} is identical to that of the SM, with the same gauge quantum numbers. The three generations of the left-chiral $SU(2)$ doublets (both quarks and leptons) transform as triplets under $A_4$ whereas the three right-chiral $SU(2)$ singlets transform as $\s, \spr$ and $\sppr$ under $A_4$. Three right-handed neutrinos are added to the model to generate neutrino masses. They transform as a triplet under $A_4$ but are singlets under the SM gauge group. The model contains multiple scalar fields, which are needed to generate PMNS matrix in TBM form. These fields consist of $\phi_0$, which is an $SU(2)$ doublet but an $A_4$ singlet, $\phi_i$~$(i = 1,2,3)$, which are $SU(2)$ doublets and form an $A_4$ triplet and $\chi_i$~$(i=1,2,3)$, which are $SU(2)$ singlets but form an $A_4$ triplet. The complete field content of the model, along with their gauge and $A_4$ quantum numbers, are summarised in table~\ref{T1}.

\begin{table}[h]
    \centering
       \begin{tabular}{|c|c|c|}
    \hline
        Fields & gauge quantum number & $A_4$ quantum number \\
        \hline
        $\left(
\begin{array}{c}
u_{iL}\\
d_{iL}
\end{array}
\right) $ & $\left( 3,2,1/3 \right)$ & \3\\          \hline
$u_{1R} \oplus u_{2R} \oplus u_{3R}$ & $\left( 3,1,4/3\right)$ &$\left(\s \oplus \spr \oplus \sppr \right) $ \\
\hline
$d_{1R} \oplus d_{2R} \oplus d_{3R}$&$\left( 3,1,-2/3
\right)$&$\left(\s \oplus \spr \oplus \sppr \right)$\\
          \hline
       $    \left(
\begin{array}{c}
\nu_{iL}\\
\bar{l}_{iL}
\end{array}
\right)$ & $(1,2,-1)$ &\3\\
\hline
$\ell_{1R} \oplus \ell_{2R} \oplus \ell_{3R}$ & $\left( 1,1,-2 \right)$& $\left(\s
\oplus \spr \oplus \sppr \right)$\\
\hline \hline
$\nu_R$ & $\left( 1,1,0 \right)$& \3 \\ \hline
$\phi_0$ &$\left( 1,2,1 \right)$ & \s \\
\hline
$\phi_i$ &$\left( 1,2,1 \right)$ &\3\\
\hline
$\chi_i $&$\left( 1,1,0 \right)$ & \3 \\
\hline
    \end{tabular}
    \caption{The field content of the model and their gauge ($SU(3) \otimes SU(2) \otimes U(1)$) and $A_4$ quantum numbers.}
    \label{T1}
\end{table}

\subsection{The Yukawa Lagrangian and Fermion Masses}\label{Lagrangian}
 The complete Yukawa Lagrangian of this model, incorporating both the gauge-invariant and $A_4$-symmetric terms, as well as the bare Majorana mass terms, can be expressed as ~\cite{He:2006dk, Grimus:2011fk}:
 \be\label{YLag}
 \mathcal{L}_{Yuk}
 \es -\Big[h_{1d} \,(\overline{Q}_{1L}\,\phi_1+\overline{Q}_{2L}\,\phi_2+\overline{Q}_{3L}\,\phi_3)\,d_{1R} 
+h_{2d}\,(\overline{Q}_{1L}\,\phi_1+\omega^2\overline{Q}_{2L}\,\phi_2+\omega\,\overline{Q}_{3L}\,\phi_3)\,d_{2R} \nn
  &&+h_{3d} \,(\overline{Q}_{1L}\,\phi_1+\omega\,\overline{Q}_{2L}\,\phi_2+\omega^2\,\overline{Q}_{3L}\,\phi_3)\,d_{3R}
 +h_{1u} \,(\overline{Q}_{1L}\,\tilde{\phi}_1+\overline{Q}_{2L}\,\tilde{\phi}_2+\overline{Q}_{3L}\,\tilde{\phi}_3)\,u_{1R} \nn
 &&+h_{2u} \,(\overline{Q}_{1L}\,\tilde{\phi}_1+\omega^2\,\overline{Q}_{2L}\,\tilde{\phi}_2+\omega\,
 \overline{Q}_{3L}\,\tilde{\phi}_3)\,u_{2R} 
 +h_{3u} \,(\overline{Q}_{1L}\,\tilde{\phi}_1+\omega\,\overline{Q}_{2L}\,\tilde{\phi}_2+\omega^2\,\overline{Q}_{3L}\,\tilde{\phi}_3)\,u_{3R}+ h.c. \Big]\nn
 &&
 -\Big[ h_{1\ell} (\overline{D}_{1L}\phi_1+\overline{D}_{2L}\phi_2+\overline{D}_{3L}\phi_3)\ell_{1R} 
 +h_{2\ell}(\overline{D}_{1L}\phi_1+\omega^2\overline{D}_{2L}\phi_2+\omega\overline{D}_{3L}\phi_3)\ell_{2R} \nn
 &&+ h_{3\ell}(\overline{D}_{1L}\phi_1+\omega\overline{D}_{2L}\phi_2+\omega^2\overline{D}_{3L}\phi_3)\ell_{3R}
 +h_0(\overline{D}_{1L}\,\nu_{1R}+\overline{D}_{2L}\,\nu_{2R}+\overline{D}_{3L}\,\nu_{3R})\tilde{\phi_0} + h.c. \Big]\nn
&&+ \frac1{2} \Big[M(\nu_{1R}^T\, C^{-1}\, \nu_{1R}+\nu_{2R}^T\, C^{-1}\, \nu_{2R}+\nu_{3R}^T\, C^{-1}\, \nu_{3R})+h.c.\Big]\nn
&&+ \frac1{2} \Big[h_\chi\, ((\chi_1 (\nu_{2R}^T\, C^{-1}\, \nu_{3R}+\nu_{3R}^T\, C^{-1}\, \nu_{2R})+
\chi_2 (\nu_{3R}^T\, C^{-1}\, \nu_{1R}+\nu_{1R}^T\, C^{-1}\, \nu_{3R})\nn
&&+ \chi_3 (\nu_{1R}^T\, C^{-1}\, \nu_{2R}+\nu_{2R}^T\, C^{-1}\, \nu_{1R}))+h.c.\Big],
 \ee
where $\tilde{\phi}_i = i \sigma_2 \phi^*_i$, $\tilde{\phi}_0 = i \sigma_2 \phi_0^*$ and   $\omega$ is the cube root of unity. $Q_{iL} (D_{iL})$ denote the left chiral $SU(2)$ doublets of quarks (leptons) with $i=1,2,3.$
 
It can be shown that the Higgs potential has a minimum
for the following vacuum expectation values (VEVs) of the scalar fields \cite{He:2006dk}:
\begin{equation}\label{vevs}
    \langle \phi_0 \rangle = \left( \begin{array}{c} 0 \\ v_0 
    \end{array} \right) \,\,; \,\, \langle \phi_i \rangle 
    = \left( \begin{array}{c} 0 \\ v 
    \end{array} \right) \,\, ; \,\,
    \langle \chi_1 \rangle = 0 = \langle \chi_3 \rangle \,\, {\rm and} \,\, \langle \chi_2 \rangle = w.
\end{equation}
That is, all three members of the $A_4$ scalar triplet $\phi$ have the same VEV $v$. With these VEVs, we obtain fermion masses of the following form
\begin{equation}
    -\bar{f}_L M_f f_R - \bar{\nu}_L \,M_D \,\nu_R
    +  \frac{1}{2} \nu_R^T C^{-1} \,M_R \,\nu_R + h.c.
\end{equation}
The charged fermion mass matrix $M_f$ can be diagonalised by making the following transformations on the left and right-chiral components
\begin{equation}
    f_{m L} = U_L^\dagger \, f_L \, \, {\rm and} \, \,
    f_{m R} = U_R\,  f_R, 
\end{equation}
 where
$U_L = U_\omega^\dagger $, $U_R = I$, and 
\,\,
\begin{equation} 
U_\omega = \frac{1}{\sqrt{3}}\left(
\begin{array}{c c c}
1 & ~~1~~ & 1\\[2ex]
1 & \omega & \omega^2\\[2ex]
1 & \omega^2 & \omega
\end{array}
\right).
\end{equation}
The Yukawa couplings of the charged fermions are simply related to their mass eigenvalues as 
\be
h_{i f} =\frac{ m_{i f}}{\sqrt{3} v}
\ee
where $i$ is the generational index. Since $m_{3 f} >>  m_{2 f} >> m_{1 f}$ that implies  $h_{3 f} >> h_{2 f}>> h_{1 f}$. The two VEVs, $v_0$ and $v$ are independent, but they satisfy the constraint $v_0^2 + 3 \, v^2 = v_{\rm EW}^2$, where $v_{\rm EW} = 174$ GeV is the VEV of the SM Higgs doublet. To reduce the number of undetermined parameters, we assume $v_0 = v$, which implies $v = v_{\rm EW}/2$.

The Dirac mass matrix for the neutrinos $M_D$ is constructed to be diagonal and the real symmetric Majorana mass matrix $M_R$ is diagonalised by an orthogonal matrix $O$. The heavy neutrino mass eigenvalues are
\begin{equation}
    (M-M^\prime), \,\, M, \,\, (M+M^\prime)
    \end{equation}
where $M$ and $M^\prime$ are the diagonal and non-diagonal elements of $M_R$ respectively. The seesaw mechanism leads to the light neutrino masses 
\begin{equation}
\frac{m_D^2}{M-M^\prime}, \,  \frac{m_D^2}{M}, \,\,  \frac{m_D^2}{M+M^\prime},
\end{equation}
where $m_D$ is the common Dirac mass of all three neutrino flavours. The PMNS matrix is now given by 
\begin{equation}
    V_{PMNS} = U_\omega^\dagger O = V_{TBM} U_{ph},
\end{equation}
where $V_{TBM}$ is the TBM form of the PMNS matrix and $U_{ph}$ is a diagonal phase matrix. 

The generation of light neutrino masses, through type-I seesaw, leads to light-heavy mixing given by $m_D M^{-1}$. Such mixing gives rise to small deviations of the PMNS matrix from unitarity. We estimate this 
deviation by comparing the expressions for light neutrino masses with data. The constraints on neutrino mass-squared differences from the oscillation data \cite{Esteban:2024eli} and that on the sum of the light neutrino masses from the cosmological data \cite{Planck:2018vyg} fix the two ratios
$m_D^2/M$ and $M^\prime/M$. An important motivation for the seesaw mechanism is to have the Dirac mass of neutrinos to be of the order of the electron mass ~\cite{Minkowski:1977sc}. To achieve this, we set $M= 1$~TeV as demonstrated in fig.~\ref{massplot} and obtain
\begin{itemize}
    \item $m_D \sim 0.1$ MeV and $M^\prime \approx 0.9\,M$ for the case of NH and
    \item $m_D \sim 0.6$ MeV and $M^\prime \approx -1.99\,M$ for the 
    case of IH.
\end{itemize}
The calculations of these parameters are given in the Appendix \ref{neutrinomass}.
Since $m_D M^{-1}$ is negligibly small (${\cal O} \sim 10^{-7}$), the CLFV amplitudes, arising due to non-unitarity of PMNS matrix, will also be negligibly small \cite{Dinh:2012bp, Awasthi:2024nvi}.

\section{The Higgs Potential and the mass eigenbasis of neutral scalars}\label{ScalarPot}
In this section, we derive the expressions for the mass eigenstates of the neutral scalars of the model. The most general scalar potential is given by  

\begin{equation}\label{genpot}
V= V(\phi_i)+ V(\phi_0) + V(\chi_i)+V(\phi_i,\chi_i)+V(\phi_i,\phi_0)+V(\phi_0,\chi_i).
\end{equation}
There is no potential term with all three fields $V(\phi_i,\phi_0,\chi_i)$ as discussed in Ref~\cite{He:2006dk}. 
The expressions for each of the terms in eq.~\ref{genpot} are given by  
\be\label{phii}
V(\phi_i) &=& \mu^2_1 (\PdagP)_{\s} +\lambda_1
(\PdagP)_{\s}(\PdagP)_{\s} + \lambda_2
(\PdagP)_{\s'}(\PdagP)_{\s''}\nonumber\\
\ps \lambda_3 (\PdagP)_{\3s}(\Pdag\phi_i)_{\3s} + \lambda_4 (\Pdag\phi_i)_{\3a}(\PdagP)_{\3a}+ i \lambda_5 (\PdagP)_{\3s}(\Pdag \phi_i)_{\3a},\\
\label{phi0}
V(\phi_0)\es \mu^2_2~ (\pdag \phi_0) + \lambda ~
(\pdag\phi_0)^2,\\ 
\label{chii}
V(\chi_i) &=& \mu^2_3 (\chi_i \chi_i)_{\s} +\eta_1 (\chi_i\chi_i\chi_i)_{\s} + \eta_2 (\chi_i\chi_i)_{\s} (\chi_i\chi_i)_{\s} + \eta_3 (\chi_i\chi_i)_{\s'}(\chi_i\chi_i)_{\s''}\nonumber\\& +& \eta_4 (\chi_i\chi_i)_{\3}(\chi_i\chi_i)_{\3},\\
\label{phiichii}
V(\phi_i,\chi_i) &=& \delta_1(\PdagP)_{\3s} \chi_i + i \delta_2 (\PdagP)_{\3a}\chi_i+
\delta_3 (\Pdag\phi_i)_{\s}(\chi_i\chi_i)_{\s}+\delta_4 (\Pdag\phi_i)_{\s'}(\chi_i\chi_i)_{\s''} \nonumber\\
&+&\delta_4^*
(\Pdag\phi_i)_{\s''}(\chi_i\chi_i)_{\s'}+\delta_5 (\Pdag\phi_i)_{\3s}(\chi_i\chi_i)_{\3} +i \delta_6 (\Pdag\phi_i)_{\3a}(\chi_i\chi_i)_{\3},\\
\label{phiiphi0}
V(\phi_i, \phi_0) \es \rho_1 (\Pdag\phi_i)_{\s} (\pdag \phi_0)+ \rho_2 (\Pdag \phi_0)(\pdag \phi_i) + \rho_3 (\Pdag \phi_0)(\Pdag \phi_0)+ \rho_3^* (\pdag \phi_i)(\pdag \phi_i),\\
\label{phi0chii}
V(\phi_0, \chi_i) \es \sigma (\pdag \phi_0) (\chi_i\chi_i)_{\s}.
\ee

In our previous work, we considered a simplified version of the above potential. We ignored all the terms containing the fields $\chi_i$. In addition, we retained only the terms of the form $(\PdagP)_{\s}$ in eqs.~\ref{phii} and \ref{phiiphi0}. In the present work, we consider the full scalar potential with the assumption that it is CP conserving. Hence, we set $\lambda_5=\delta_6=\delta_2=0$.
Minimisation of the potential leads to the following additional constraints: the trilinear coupling term $\delta_1=0$,  
$\rho_3$ is real and $\delta_4=\delta_4^*=0$.

The potential contains the following set of neutral scalars:
$\Re  \phi_\alpha =( \Re \phi_0, \Re \phi_i)$, 
$\Im  \phi_\alpha =( \Im \phi_0, \Im \phi_i)$ and $\chi_i$. After minimising the potential, we obtain the mass-squared matrices for the real and imaginary components of the scalars.
The mass-squared matrix for the real scalars is a $7 \times 7 $ matrix of the form
\begin{equation}
    M^2_{7\times 7}=\left[ 
    \begin{array}{cc}
      \mathcal{M}^2_{\Re \phi_\alpha \Re \phi_\beta}   & \mathcal{M}^2_{\Re \phi_\alpha \chi_i}  \\
       (\mathcal{M}^2_{\Re \phi_\alpha \chi_i})^T   &  \mathcal{M}^2_{\chi_i \chi_j} 
    \end{array}\right], 
\end{equation}
where 
\be\label{MRphi}
\mathcal{M}^2_{\Re \phi_\alpha \Re \phi_\beta} \es
\left(
\begin{array}{cccc}
A & B & B & B \\
 B & C& D & D \\
 B & D & C& D \\
 B & D & D & C\\
\end{array}
\right),
\ee
where
$A = 2 \lambda  v_0^2,~B = (\rho_1+\rho_2+2 \rho_3) v v_0,~ C= 2 (\lambda_1+\lambda_2) v^2 $ and $D=(2 \lambda_1-\lambda_2+4 \lambda_3) v^2.$
\be\label{Mphichi}
 \mathcal{M}^2_{\Re \phi_\alpha \chi_i}  \es \left(
\begin{array}{ccc}
 0 & ~\sigma v_0 w & 0 \\
\delta_5 v w &~ \delta_3 v w  & 0 \\
\delta_5 v w &~ \delta_3 v w &\delta_5 v w \\
 0 &~\delta_3 v w  &\delta_5 v w \\
\end{array}\right),\\
\label{Mchi}
 \mathcal{M}^2_{\chi_i \chi_j} \es \left(
\begin{array}{ccc}
 \frac{1}{2}  (4 \eta_4-3 \eta_3) w^2 &\delta_5 v^2 &\delta_5 v^2 \\
\delta_5 v^2 & 2 (\eta_2+\eta_3)w^2 &\delta_5 v^2 \\
\delta_5 v^2 &\delta_5 v^2 & \frac{1}{2} (4 \eta_4-3 \eta_3)w^2 \\
\end{array}
\right).
\ee
The mass-squared matrix for the imaginary component is given by 
\be \label{MIphi}
\mathcal{M}^2_{\Im \phi_\alpha \Im\phi_\beta} \es \left(
\begin{array}{cccc}
A' & B' & B' & B' \\
 B' & C'& D' & D' \\
 B' & D' & C'& D' \\
 B' & D' & D' & C'\\
\end{array}
\right),
\ee
where
$A' =-6 \rho_3 v^2 ,~B' =  2 \rho_3 v v_0,~ C'= -4 (\lambda_3+\lambda_4) v^2-2 \rho_3 v_0^2$ and $D' = 2(\lambda_3+\lambda_4) v^2.$
The three matrices, $\mathcal{M}^2_{\Re \phi_\alpha \Re\phi_\beta} $, $\mathcal{M}^2_{\Im \phi_\alpha \Im\phi_\beta} $ and $\mathcal{M}^2_{\chi_i \chi_j} $ are real symmetric matrices and are diagonalized by appropriate orthogonal transformations.  

We note that the two matrices,  $\mathcal{M}^2_{\Re \phi_\alpha \Re\phi_\beta} $ and $\mathcal{M}^2_{\Im \phi_\alpha \Im\phi_\beta}$, have the same structure as seen in eq.~\ref{MRphi} and eq.~\ref{MIphi}. Therefore, their diagonalizing matrices will also have the same structure. We denote the eigenstates of $\mathcal{M}^2_{\Re \phi_\alpha \Re\phi_\beta} $ and $\mathcal{M}^2_{\Im \phi_\alpha \Im\phi_\beta}$ by $\Re \Phi_\alpha$ and $\Im \Phi_\alpha$ respectively. 
The $A_4$ eigenbasis of SU(2) doublet scalars $\Re \phi_\alpha$
can be expressed in terms of $\Re\Phi_\beta$ through the orthogonal transformation $\mathcal{O}_R$ 
\be
\Re\phi_\alpha \es (\mathcal{O}_R)_{\alpha \beta} \Re\Phi_\beta, 
\ee
where 
\be\label{UnitaryHRexp}
\mathcal{O}_R=  
\left(
\begin{array}{cccc}
\frac{a_{1R}}{\sqrt{3+a_{1R}^2}} & \frac{a_{2R}}{\sqrt{3+a_{2R}^2}} & 0 & 0 \\[2ex]
\frac{1}{\sqrt{3+a_{1R}^2}} & \frac{1}{\sqrt{3+a_{2R}^2}} & -\frac{1}{\sqrt{2}}   &  \frac{1}{\sqrt{6}} \\[2ex]
\frac{1}{\sqrt{3+a_{1R}^2}} & \frac{1}{\sqrt{3+a_{2R}^2}} & \frac{1}{\sqrt{2}}   &  \frac{1}{\sqrt{6}} \\[2ex]
\frac{1}{\sqrt{3+a_{1R}^2}}& \frac{1}{\sqrt{3+a_{2R}^2}} & 0 & \frac{-2}{\sqrt{6}}
\end{array}
\right).
\ee
The quantities $a_{1R}$ and $a_{2R}$ are given by
\be \label{a1R} a_{1R, 2R} \es \frac{X \mp\sqrt{X^2 +3 Y^2}}{Y},
\ee
where $ X =  (\lambda +\rho_1+\rho_2+2 \rho_3) v_0^2-(3 \lambda_1+4 \lambda_3+3 \rho_3) v^2, \,\,\,    Y = (\rho_1+\rho_2+2 \rho_3) v v_0$.

We observe that $a_{1R} a_{2R} =-3$, which ensures the orthogonality of the first two columns.
The $A_4$ eigenbasis of SU(2) doublet scalars $\Im \phi_\alpha$
can be expressed in terms of $\Im\Phi_\beta$ through the orthogonal transformation $\mathcal{O}_I$ 
\be
\Im\phi_\alpha \es (\mathcal{O}_I)_{\alpha \beta} \Im\Phi_\beta, 
\ee
where 
\be\label{UnitaryHIexp}
\mathcal{O}_I=  
\left(
\begin{array}{cccc}
\frac{a_{1I}}{\sqrt{3+a_{1I}^2}} & \frac{a_{2I}}{\sqrt{3+a_{2I}^2}} & 0 & 0 \\[2ex]
\frac{1}{\sqrt{3+a_{1I}^2}} & \frac{1}{\sqrt{3+a_{2I}^2}} & -\frac{1}{\sqrt{2}}   &  \frac{1}{\sqrt{6}} \\[2ex]
\frac{1}{\sqrt{3+a_{1I}^2}} & \frac{1}{\sqrt{3+a_{2I}^2}} & \frac{1}{\sqrt{2}}   &  \frac{1}{\sqrt{6}} \\[2ex]
\frac{1}{\sqrt{3+a_{1I}^2}}& \frac{1}{\sqrt{3+a_{2I}^2}} & 0 & \frac{-2}{\sqrt{6}}
\end{array}
\right),
\ee
and $a_{1I}=v_0/v$ and $a_{2I}=-3v/v_0$. We note that the two diagonalizing matrices $\mathcal{O}_R$ and $\mathcal{O}_I$ have similar structure. For this reason, the couplings of charged fermions to $\Re \Phi_\alpha$ and $\Im \Phi_\alpha$ will also be similar. The eigenvalues of $\mathcal{M}^2_{\Re \phi_\alpha \Re \phi_\beta} $, corresponding to eigenvectors $\Re\Phi_2$ and $\Re\Phi_3$ are degenerate. The matrix  $\mathcal{M}^2_{\Im \phi_\alpha \Im \phi_\beta}$ has one zero eigenvalue, which corresponds to Goldstone boson coupling to $Z^0$, and two degenerate eigenvalues corresponding to eigenvectors $\Im\Phi_2$ and $\Im\Phi_3$.
From the expressions for $\mathcal{O}_R$ and $\mathcal{O}_I$, we note that the last two eigenvectors, whose eigenvalues are degenerate, are identical. Thus, any arbitrary linear combination of these vectors will also have degenerate eigenvalues.

Since we assume the scalar potential to be CP-conserving, there will be no mixing between the real and the imaginary components of the scalars. Hence, $\Im \Phi_\alpha$ are mass eigenstates. However, there will be mixing between $\Re \Phi_\alpha$ and $\chi_i$
due to $\mathcal{M}^2_{\Re \phi_\alpha \chi_i}$. 
In Appendix \ref{NSmassmatrix}, we will show that this mixing is constrained to be extremely small by the tiny value of $K^0-\bar{K}^0$ mixing. Hence, we take $\Re \Phi_\alpha$ also to be the mass eigenstates. Since $\chi_i$ does not couple to charged fermions, the eigenvalue problem of $\mathcal{M}^2_{\chi_i \chi_j}$ is not relevant for our work. 

We first consider the couplings of the complex neutral scalars of the doublets $\phi_0$ and $\phi_i$ to down-type quarks in their $A_4$ eigenbasis. The corresponding Yukawa Lagrangian is 
\be\label{YCiFME}
\mathcal{L}^d_{Yuk} \es-\frac{h_{1d}}{\sqrt{3}}\Big[
\left(\bar{d}_L+\bar{s}_L+\bar{b}_L\right)\phi_{1}+\left(\bar{d}_L+\omega\bar{s}_L+\omega^2\bar{b}_L\right)\phi_{2}+\left(\bar{d}_L+\omega^2 \bar{s}_L+\omega \bar{b}_L\right)\phi_{3}\Big]d_R\nn
\ms \frac{h_{2d}}{\sqrt{3}}\Big[
\left(\bar{d}_L+\bar{s}_L+\bar{b}_L\right)\phi_{1}
+\omega^2\left(\bar{d}_L+\omega\bar{s}_L+\omega^2\bar{b}_L\right)\phi_{2}
+\omega\left(\bar{d}_L+\omega^2 \bar{s}_L+\omega \bar{b}_L\right)\phi_{3}\Big]s_R\nn
\ms \frac{h_{3d}}{\sqrt{3}}\Big[
\left(\bar{d}_L+\bar{s}_L+\bar{b}_L\right)\phi_{1}
+\omega\left(\bar{d}_L+\omega\bar{s}_L+\omega^2\bar{b}_L\right)\phi_{2}
+\omega^2\left(\bar{d}_L+\omega^2 \bar{s}_L+\omega \bar{b}_L\right)\phi_{3}\Big]b_R
\nn
\ps h.c.
\ee
Since the diagonalizing matrices $\mathcal{O}_R$ and $\mathcal{O}_I$ have the same structure, the complex fields $\phi_\alpha$ can be transformed to the mass eigenbasis $\Phi_\alpha$ by the following generic orthogonal matrix $\mathcal{O}$ given by
\be\label{UnitaryHexp}
\mathcal{O}=  
\left(
\begin{array}{cccc}
\frac{a_{1}}{\sqrt{3+a_{1}^2}} & \frac{a_{2}}{\sqrt{3+a_{2}^2}} & 0 & 0 \\[2ex]
\frac{1}{\sqrt{3+a_{1}^2}} & \frac{1}{\sqrt{3+a_{2}^2}} & -\frac{1}{\sqrt{2}}   &  \frac{1}{\sqrt{6}} \\[2ex]
\frac{1}{\sqrt{3+a_{1}^2}} & \frac{1}{\sqrt{3+a_{2}^2}} & \frac{1}{\sqrt{2}}   &  \frac{1}{\sqrt{6}} \\[2ex]
\frac{1}{\sqrt{3+a_{1}^2}}& \frac{1}{\sqrt{3+a_{2}^2}} & 0 & \frac{-2}{\sqrt{6}}
\end{array}
\right),
\ee
The Yukawa Lagrangian, in the mass eigenbasis of the down-type quarks and the complex fields $\Phi_\alpha$, is 

\be\label{YCwh2h3}
\mathcal{L}_{Yuk}^{d}\es
- \frac{h_{1d}}{\sqrt{3}}\Bigg[\left(\bar{d}_L+\bar{s}_L+\bar{b}_L\right)\left(\frac{1}{\sqrt{3+a_1^2}}\Phi_{0}+\frac{1}{\sqrt{3+a_2^2}}\Phi_{1}-\frac{1}{\sqrt{2}}\Phi_{2}+\frac{1}{\sqrt{6}}\Phi_{3}\right)\nonumber \\ 
\ps \left(\bar{d}_L+\omega\bar{s}_L+\omega^2\,\bar{b}_L\right)\left( \frac{1}{\sqrt{3+a_1^2}}\Phi_{0}+\frac{1}{\sqrt{3+a_2^2}}\Phi_{1}+\frac{1}{\sqrt{2}}\Phi_{2}+\frac{1}{\sqrt{6}}\Phi_{3} \right)\nonumber \\ 
\ps\left(\,\bar{d}_L+\omega^2\bar{s}_L+\omega\, \bar{b}_L\right)\left( \frac{1}{\sqrt{3+a_1^2}}\Phi_{0}+\frac{1}{\sqrt{3+a_2^2}}\Phi_{1}-\frac{2}{\sqrt{6}}\Phi_{3} \right)\Bigg]d_R\nonumber \\
\ms \frac{h_{2d}}{\sqrt{3}}\Bigg[\left(\bar{d}_L+\bar{s}_L+\bar{b}_L\right)\left(\frac{1}{\sqrt{3+a_1^2}}\Phi_{0}+\frac{1}{\sqrt{3+a_2^2}}\Phi_{1}-\frac{1}{\sqrt{2}}\Phi_{2}+\frac{1}{\sqrt{6}}\Phi_{3}\right)\nonumber \\ 
\ps \left(\omega^2 \bar{d}_L+\bar{s}_L+\omega\,\bar{b}_L\right)\left( \frac{1}{\sqrt{3+a_1^2}}\Phi_{0}+\frac{1}{\sqrt{3+a_2^2}}\Phi_{1}+\frac{1}{\sqrt{2}}\Phi_{2}+\frac{1}{\sqrt{6}}\Phi_{3} \right)\nonumber \\ 
\ps\left(\omega\,\bar{d}_L+\bar{s}_L+\omega^2\, \bar{b}_L\right)\left( \frac{1}{\sqrt{3+a_1^2}}\Phi_{0}+\frac{1}{\sqrt{3+a_2^2}}\Phi_{1}-\frac{2}{\sqrt{6}}\Phi_{3} \right)\Bigg]s_R\nonumber \\
\ms \frac{h_{3d}}{\sqrt{3}}\Bigg[\left(\bar{d}_L+\bar{s}_L+\bar{b}_L\right)\left(\frac{1}{\sqrt{3+a_1^2}}\Phi_{0}+\frac{1}{\sqrt{3+a_2^2}}\Phi_{1}-\frac{1}{\sqrt{2}}\Phi_{2}+\frac{1}{\sqrt{6}}\Phi_{3}\right)\nonumber \\ 
\ps \left(\omega \bar{d}_L+\omega^2\,\bar{s}_L+\,\bar{b}_L\right)\left( \frac{1}{\sqrt{3+a_1^2}}\Phi_{0}+\frac{1}{\sqrt{3+a_2^2}}\Phi_{1}+\frac{1}{\sqrt{2}}\Phi_{2}+\frac{1}{\sqrt{6}}\Phi_{3} \right)\nonumber \\ 
\ps\left(\omega^2\bar{d}_L+\omega\bar{s}_L+ \bar{b}_L\right)\left( \frac{1}{\sqrt{3+a_1^2}}\Phi_{0}+\frac{1}{\sqrt{3+a_2^2}}\Phi_{1}-\frac{2}{\sqrt{6}}\Phi_{3} \right)\Bigg]b_R+h.c.
\ee
From eq.~\ref{YCwh2h3}, we note that the fields $\Phi_0$ and $\Phi_1$ have purely diagonal couplings while the fields $\Phi_2$ and $\Phi_3$ have purely off-diagonal couplings.  
Hence, $\Phi_2$ and $\Phi_3$ can mediate flavour-changing interactions at tree-level. 

The flavour changing couplings between the first two generations of the d-type quarks, given in eq.~\ref{YCwh2h3}, in terms of the mass eigenstates $\Re \Phi_2, \Re \Phi_3, \Im \Phi_2$ and $\Im \Phi_3$ are
\be\label{YCrealimg}
\mathcal{L}_{Yuk}^{d}\es
- \frac{h_{1d}}{\sqrt{12}}\bar{s}_L\left((-1+\omega)\Re\Phi_{2}-\sqrt{3}\omega^2\Re\Phi_{3}\right)d_R
-\frac{h_{2d}}{\sqrt{12}}\bar{d}_L\left((-1+\omega^2)\Re\Phi_{2}-\sqrt{3}\omega\Re\Phi_{3}\right)
s_R\nonumber\\
\ms i\frac{h_{1d}}{\sqrt{12}}\bar{s}_L\left((-1+\omega)\Im\Phi_{2}-\sqrt{3}\omega^2\Im\Phi_{3}\right)d_R
-i\frac{h_{2d}}{\sqrt{12}}\bar{d}_L\left((-1+\omega^2)\Im\Phi_{2}-\sqrt{3}\omega\Im\Phi_{3}\right)
s_R\nn
\ps h.c.
\ee
From these couplings, we calculate the effective four-fermion Hamiltonian for $K^0-\bar{K}^0$ transition \be\label{kkbarmix}
H_{eff} (K^0 \to \bar{K}^0)\es\frac{\omega}{4}\bigg(h_{1d}^2\; (\bar{s}_L d_R)^2+h_{2d}^2\; (\bar{s}_R d_L)^2\bigg)\left[ \left(\frac{1}{m_{\Re\Phi_3}^2} -\frac{1}{m_{\Re\Phi_2}^2}\right)-
\left(\frac{1}{m_{\Im\Phi_3}^2}-\frac{1}{m_{\Im\Phi_2}^2}\right) 
 \right] \nn
\ps 
2 \frac{\omega}{4} h_{1d}h_{2d} (\bar{s}_R d_L)(\bar{s}_L d_R)
\left[ \left(\frac{1}{m_{\Re\Phi_3}^2} -\frac{1}{m_{\Re\Phi_2}^2}\right)+
\left(\frac{1}{m_{\Im\Phi_3}^2}-\frac{1}{m_{\Im\Phi_2}^2}\right) 
 \right].\nn
\ee
The amplitudes arising due to the exchange of $\Re \Phi_2$ and $\Re \Phi_3$ cancel each other due to the degeneracy of the masses. The same is true for the exchange of $\Im \Phi_2$ and $\Im \Phi_3$. Hence, there is no $K^0-\bar{K}^0$ mixing due to tree-level flavour-changing couplings of $\Phi_2$ and $\Phi_3$. Similarly, there is no mixing of neutral $B$-mesons. Hence, the constraints from the neutral meson mixing are trivially satisfied in this model. 

We consider the following complex linear combinations  
\be\label{complexcomb}
\tilde{\Phi}_2 \es \frac{-\Phi_3 + i \Phi_2}{\sqrt{2}},~~~~~~\tilde{\Phi}_3 = \frac{-\Phi_3 - i \Phi_2}{\sqrt{2}}.
\ee
Degeneracy of the masses of $\Re \Phi_2$ and $\Re \Phi_3$ implies that the masses of $\Re \tilde{\Phi}_2$ and $\Re \tilde{\Phi}_3$ are also degenerate. The same statement holds for imaginary components also. The Yukawa couplings of negatively charged fermions to these linear combinations 
take a very simple form, given below
\begin{equation}\label{dqflch}
g^{ij}_2\bar{f}_{iL} f_{jR} \tilde{\Phi}_2 + g^{ij}_3\bar{f}_{iL} f_{jR} \tilde{\Phi}_3
+ \left( g^{ji}_2\right)^* \bar{f}_{iR} f_{jL} \left( \tilde{\Phi}_2 \right)^* 
+ \left( g^{ji}_3\right)^* \bar{f}_{iR} f_{jL} \left( \tilde{\Phi}_3 \right)^*.
\end{equation}
The flavour-changing couplings for neutral scalars and up-type quarks have a slightly different structure, so we give its form
\begin{equation}\label{uqflch}
g^{ij}_2 \bar{u}_{iL} u_{jR} (\tilde{\Phi}_2)^* + g^{ij}_3\bar{u}_{iL} u_{jR} (\tilde{\Phi}_3)^*
+ \left( g^{ji}_2 \right)^* \bar{u}_{iR} u_{jL}  \tilde{\Phi}_2 
+\left(g^{ji}_3 \right)^* \bar{u}_{iR} u_{jL} \tilde{\Phi}_3 .
\end{equation}
We find that, if $g^{ij}_2 =0$ then $g^{ij}_3 \neq 0$ and vice-versa. The set of flavour changing couplings of charged fermions with $T_3= -1/2$ and $T_3= 1/2$ are listed in table \ref{yucup}.

\begin{table}[h]
    \centering
\begin{tabular}{|c|c|c|}
\hline
  & &\\[-2.3ex]
fermions & scalar field $\tilde{\Phi}_2$ & scalar field $\tilde{\Phi}_3$ \\
  \hline
  & $g_2^{\mu e}=h_{1\ell}\, \omega^2 $ & $g_3^{\tau e}=h_{1\ell}\,
 \omega $ \\
  
 leptons &$g_2^{\tau \mu}=h_{2\ell}\, \omega^2$ & $g_3^{e \mu}=h_{2\ell}\,
 \omega$ \\
 &$g_2^{e \tau}=h_{3\ell}\, \omega^2$ & $g_3^{\mu \tau}=h_{3\ell}\,
 \omega $ \\
 & $g_2^{e \mu}=g_2^{\mu \tau}=g_2^{\tau e}=0$ & $g_3^{e \tau}=g_3^{\mu e}=g_3^{\tau \mu}=0$
   \\
  \hline
  &$g_2^{sd}=h_{1d}\,\omega^2 $ & 
  $g_3^{bd}=h_{1d}\,\omega$\\
  down-type &$g_2^{bs}=h_{2d}\,\omega^2$ & 
  $g_3^{ds}=h_{2d}\,\omega$
   \\
  &$g_2^{db}=h_{3d}\,\omega^2$ & $g_3^{sb}=h_{3d}\,\omega $  \\
  &$g_2^{ds}=g_2^{sb}=g_2^{db}=0$& $g_3^{db}=g_3^{sd}=g_3^{bs}=0$\\
  \hline
    \hline
    & &\\[-2.3ex]
    fermions & scalar field $\tilde{\Phi}_2 ^{*}$ & scalar field $\tilde{\Phi}_3 ^{*}$ \\  \hline
  &$g^{tu}_2=h_{1u}\,\omega$& $g^{c u}_3=h_{1u}\,\omega^2$ \\
  up-type & $g^{uc}_2=h_{2u}\,\omega$  &$g^{tc}_3=h_{2u}\,\omega^2 $    \\
   & $g^{ct}_2=h_{3u}\,\omega$  &  $g_3^{ut}=h_{3u}\,\omega^2$\\
    & $g^{ut}_2=g^{cu}_2=g^{tc}_2=0$  &  $g_3^{uc}=g_3^{ct}=g_3^{tu}=0$\\
  \hline
  \end{tabular}\\
      \caption{Flavour changing couplings of the charged fermions with $T_3=-1/2 (1/2)$ to the neutral scalar fields $\tilde{\Phi}_{2/3}(\tilde{\Phi}^*_{2/3})$.}
    \label{yucup}
\end{table}
Expressing the complex scalar field of eq.~\ref{dqflch} and eq.~\ref{uqflch} in terms of their real and imaginary components, we get
\be \label{dqfchrealimg}
&&\frac{1}{\sqrt{2}}\left[ \left(g_2^{ij}\right) \bar{f}_{iL} f_{jR} \left(\Re \tilde{\Phi}_2 +i \Im \tilde{\Phi}_2\right) + \left( g_3^{ij} \right)\bar{f}_{iL} f_{jR} \left(\Re \tilde{\Phi}_3 +i \Im \tilde{\Phi}_3\right) \right.\nonumber\\ &&
\left.+\left( g_2^{ji} \right)^* \bar{f}_{iR} f_{jL} \left(\Re \tilde{\Phi}_2 -i \Im \tilde{\Phi}_2\right) 
+ \left( g_3^{ji} \right)^* \bar{f}_{iR} f_{jL} \left(\Re \tilde{\Phi}_3 - i\Im \tilde{\Phi}_3\right) \right],
\\[3ex]
\label{uqfchrealimg}
&&\frac{1}{\sqrt{2}}\left[ \left(g^{ij}_2\right) \bar{u}_{iL} u_{jR} \left(\Re \tilde{\Phi}_2 -i \Im \tilde{\Phi}_2\right)  + \left(g^{ij}_3\right)\bar{u}_{iL} u_{jR} \left(\Re \tilde{\Phi}_3 -i \Im \tilde{\Phi}_3\right) \right.\nonumber\\ &&
\left.
+\left( g^{ji}_2 \right)^* \bar{u}_{iR} u_{jL}  \left(\Re \tilde{\Phi}_2 +i \Im \tilde{\Phi}_2\right)  
+\left(g^{ji}_3 \right)^* \bar{u}_{iR} u_{jL} \left(\Re \tilde{\Phi}_3 +i \Im \tilde{\Phi}_3 \right) 
\right].
\ee
\section{Charged Lepton Flavour Violating transitions}\label{CLFVdecays}
\subsection{CLFV decays of down-type quarks}
First, we do an explicit calculation of the tree-level amplitude of the decay $K^0 \to \mu^+ e^-$. From the couplings listed in table \ref{yucup}, we note that this transition can arise due to 
the exchange of $\tilde{\Phi}_2$ as well as the exchange of $\tilde{\Phi}_3$. However, the coefficient of $\tilde{\Phi}_3$ exchange term ($h_{2d} h_{2\ell}$) is much larger than that of $\tilde{\Phi}_2$ exchange term ($h_{1d} h_{1\ell}$). Due to the hierarchy of Yukawa couplings, we consider only the amplitude corresponding to $\tilde{\Phi}_3$ exchange.
\begin{figure}[h]
    \centering
    \includegraphics[scale=0.5]{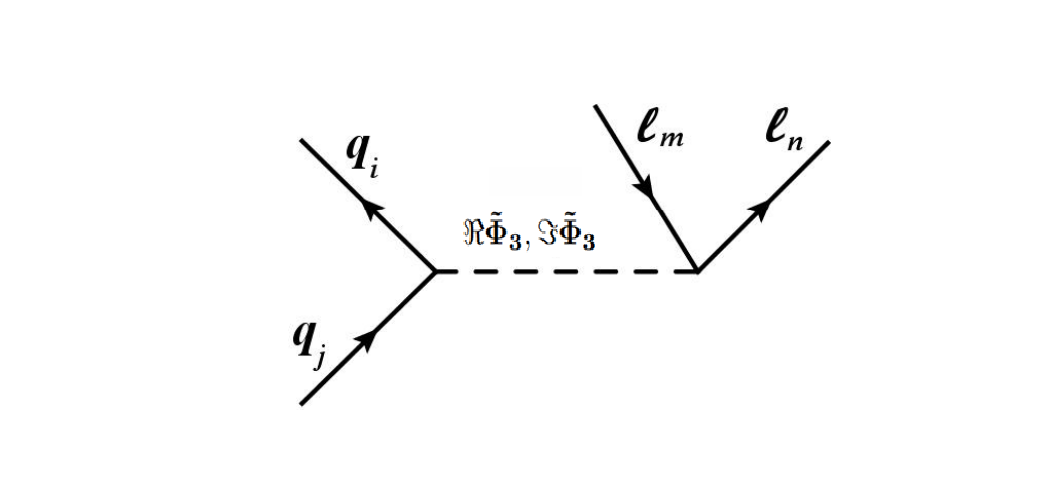}
        \caption{Tree-level Feynman diagram for $\bar{s} d \to \mu^+ e^-$ transition, mediated by $\Re \tilde{\Phi}_3$ and $\Im \tilde{\Phi}_3$.}
    \label{m2clfv}
    \end{figure}
The Feynman diagram for the transition $\bar{s} d \to \mu^+ e^-$ is given in fig.~\ref{m2clfv}.
From the couplings given in eq.~\ref{dqfchrealimg}, we get the four-fermion Hamiltonian for this transition to be
\begin{eqnarray}
{\cal H}_{eff} (K^0 \to \mu^+ e^-) =\frac{h_{2d}h_{2\ell}}{2}(\bar{s}_R d_L)\left[\frac{1}{m^2_{\Re\tilde{\Phi}_3}}+\frac{1}{m^2_{\Im\tilde{\Phi}_3}}\right](\bar{e}_L\mu_R).
\label{decayamp}
\end{eqnarray}
We calculate the decay width to be 
\be
\Gamma (K_L \to \mu^+ e^-)= \frac{2}{\pi}\left(\frac{m_K}{4} \right)^5 \left(\frac{m_\mu}{\sqrt{3}v}\right)^2 \left(\frac{f_K}{\sqrt{3}v}\right)^2\left[\frac{1}{m^2_{\Re\tilde{\Phi}_3}}+\frac{1}{m^2_{\Im\tilde{\Phi}_3}}\right]^2,
\ee
where we set $(1-m_\mu^2/m_K^2)\approx 1$ and $m_s/(m_s+m_d)\approx 1$. Comparing this expression to the experimental upper bound $BR(K_L\to \mu^+ e^-)< 4.7\times 10^{-12}$ \cite{BNL:1998apv}, we obtain the bound 
\be \label{Phi23masslimit}
\left[\frac{1}{m^2_{\Re\tilde{\Phi}_3}}+\frac{1}{m^2_{\Im\tilde{\Phi}_3}}\right] < 2.7\times 10^{-6}\, \rm{(GeV)^{-2}}.
\ee
Due to the degeneracy of the masses of $\tilde{\Phi}_2$ and $\tilde{\Phi}_3$, the above limit applies to $\tilde{\Phi}_2$ also. The bound in eq.~\ref{Phi23masslimit} implies that the masses of these scalars with flavour-changing couplings are $\geq 900$ GeV. This bound is consistent with the LHC bounds on the searches for LFV decays of neutral scalars, set by ATLAS \cite{ATLAS:2024vxm} and CMS \cite{CMS:2019pex}.

The allowed CLFV decays of neutral mesons into purely leptonic final states are listed in 
table~\ref{clfvcff}. Each decay receives two contributions: one mediated by $\tilde{\Phi}_2$ and the other mediated by $\tilde{\Phi}_3$. The products of Yukawa couplings for each contribution are listed in table~\ref{clfvcff}. In every case, one product is much larger than the other. The decay amplitudes can be calculated using the flavour-changing couplings listed in table~\ref{yucup}. Substituting the bound on the masses $\tilde{\Phi}_2/ \tilde{\Phi}_3$, the upper limits on their branching ratios can be calculated.
\begin{table}[h]
\centering
\begin{tabular}{|l|ll|}
\hline \hline
\multirow{2}{*}{Neutral mesons decay}     & \multicolumn{2}{l|}{Product of Yukawa couplings}                                                                                                                                       \\ \cline{2-3} 
                             & \multicolumn{1}{l|}{\begin{tabular}[c]{@{}l@{}} mediated by $\tilde{\Phi}_2$\end{tabular}} & \begin{tabular}[c]{@{}l@{}}mediated by $\tilde{\Phi}_3$ \end{tabular} \\ \hline 
$K^0 (\overline{s}d) \to \mu^+ e^-$       & \multicolumn{1}{l|}{$h_{1d} h_{1\ell}$}                                                                      & $ {\bf h_{2d} h_{2\ell}}$                                                                         \\ \hline
$B_d ^0 (\overline{b}d) \to e^+ \mu^-$    & \multicolumn{1}{l|}{${\bf h     _{3d} h_{1\ell}}$}                                                                      & $h_{1d} h_{2\ell}$                                                                         \\ \hline
$B_d ^0 (\overline{b}d) \to \mu^+ \tau^-$ & \multicolumn{1}{l|}{${\bf h_{3d} h_{2\ell}}$}                                                                      & $h_{1d} h_{3\ell}$                                                                         \\ \hline
$B_d ^0 (\overline{b}d) \to \tau^+ e^-$    & \multicolumn{1}{l|}{${\bf h_{3d} h_{3\ell}}$}                                                                       & $h_{1d} h_{1\ell}$                                                                         \\ \hline
$B_s ^0 (\overline{b}s) \to \mu^+ e^-$    & \multicolumn{1}{l|}{$h_{2d} h_{1\ell}$}                                                                      & ${\bf h_{3d} h_{2\ell}}$                                                                         \\ \hline
$B_s ^0 (\overline{b}s) \to \tau^+ \mu^-$ & \multicolumn{1}{l|}{$h_{2d} h_{2\ell}$}                                                                      & ${\bf h_{3d} h_{3\ell}}$                                                                         \\ \hline
$B_s ^0 (\overline{b}s) \to  e^+ \tau^-$  & \multicolumn{1}{l|}{${\bf h_{2d} h_{3\ell}}$}                                                                      & $h_{3d} h_{1\ell}$                                                                         \\ \hline
\end{tabular}
\caption{The product of Yukawa coupling coefficients for different CLFV neutral meson decays. The dominant product of Yukawa couplings is highlighted in bold.}
 \label{clfvcff}
\end{table}
We also note that if $M^0 (\bar{q}_i q_j) \to \ell_m^+ \ell_n^-$ is allowed then the CP conjugate mode $\overline{M}^0 
\to \ell_m^- \ell_n^+$ is also allowed. However, the decay with charge conjugated final state, $M^0 \to \ell_m^- \ell_n^+$, is not allowed. 
As in the case of vanishing $M^0 \leftrightarrow \overline{M}^0$ amplitude, this is a consequence of the $A_4$ symmetry of the model. 

We calculate the branching ratios of purely leptonic CLFV decays of neutral $B$ mesons in this model, by substituting the limits on the neutral scalar masses from eq.~\ref{Phi23masslimit}. 
The calculated upper bounds, predicted by the model, are listed in table~\ref{lepBR}, along with the present experimental upper bounds.
\begin{table}[h]
\centering
\begin{tabular}{|l|l|l|}
\hline
                                                                                                                                       \\ \cline{2-3} 
Neutral mesons decay & Calculated BR & Exp. upper bound  \\[1.1ex] \hline                          
$B_d ^0 (\overline{b}d) \to e^+ \mu^-$   & $1.6\times 10^{-15}$  &                              $1.0 \times 10^{-9}$ \cite{LHCb:2017hag}                                                \\[1.1ex] \hline
$B_d ^0 (\overline{b}d) \to \mu^+ \tau^-$  &  $5.3\times 10^{-11}$   &                          $1.4 \times 10^{-5}$   
\cite{LHCb:2019ujz} \\[1.1ex] \hline
$B_d ^0 (\overline{b}d) \to \tau^+ e^-$    &  $1.5\times 10^{-8}$  &   $1.6\times 10^{-5}$ \cite{Belle:2021rod}                                                                           \\[1.1ex] \hline
$B_s ^0 (\overline{b}s) \to \mu^+ e^-$     &  $1.0\times 10^{-10}$   &       $ 5.4\times 10^{-9}$ \cite{LHCb:2017hag}                                                                     \\[1.1ex] \hline
$B_s ^0 (\overline{b}s) \to \tau^+ \mu^-$   & $2.3\times 10^{-8}$  &     $4.2\times 10^{-5}$ \cite{LHCb:2019ujz}                                                                      \\[1.1ex] \hline
$B_s ^0 (\overline{b}s) \to  e^+ \tau^-$   & $1.2\times 10^{-11}$   &   
$1.4 \times 10^{-3}$  \cite{Belle:2023jwr}  \\[1.1ex] \hline
\end{tabular}
    \caption{Predicted branching ratios for purely leptonic CLFV decays of neutral $B$ mesons}\label{lepBR}
\end{table}

 \subsection{Semileptonic meson decays}\label{semileptonic}
In addition to purely leptonic CLFV decays, the pseudoscalar mesons can also have semileptonic CLFV decays. Because the final state involves three particles, the phase space will be smaller, while these additional modes provide an extra channel to search for CLFV in meson decays. These semileptonic decays can have either a pseudoscalar meson or a vector meson in the final state. 
We have computed the upper bounds on both types of semileptonic branching ratios of meson CLFV decays. In table~\ref{b2dllbar},
we list the calculated and experimental bounds of the decays driven by the 
$\bar{b} \to \bar{d} \,\ell_m^+ \ell_n^-$ transition. In table~\ref{b2sllbar}, we do the same for the decays driven by the $\bar{b} \to \bar{s} \, \ell_m^+ \ell_n^-$ transition. 
In computing these bounds, we have used kaon form factors from Ref~\cite{Carrasco:2016kpy} and $B$-meson data from Ref~\cite{Becirevic:2024vwy}. We also obtain an upper limit on the branching ratio for the decay $K^+ \to \pi^+ \mu^+ e^-$ as $4.9 \times 10^{-15}$, which is four orders of magnitude smaller than the current experimental limit of $2.1 \times 10^{-11}$ reported in Ref~\cite{Sher:2005sp}.
\begin{table}[h!]
\centering
\begin{tabular}{|c|c|c|c|}
\hline
& & &  \textbf{Dominant}\\
\textbf{Decay Mode} & \textbf{Pred. Limit} & \textbf{Exp. Limit} & \textbf{Yukawa Term} \\
\hline
\hline
$B^+ \rightarrow \pi^+ e^+ \mu^-$ & $3.2\times 10^{-16}$ & $ 6.4\times 10^{-3}$ \cite{Weir:1989sq} &  $h_{3d} h_{1\ell}$ \\
$B^+ \rightarrow \pi^+ \mu^+ \tau^-$ & $8.3\times 10^{-12}$ & $ 6.2\times 10^{-5}$ \cite{BaBar:2012azg} & $h_{3d} h_{2\ell}$ \\
$B^+ \rightarrow \pi^+ \tau^+ e^-$ & $2.5\times 10^{-9}$ & $ 2.0\times 10^{-5}$ \cite{BaBar:2012azg} & $h_{3d} h_{3\ell}$\\
\hline
$B^0 \rightarrow \pi^0  e^+ \mu^-$ &$1.6 \times 10^{-16}$ & $1.4 \times 10^{-7}$ \cite{BaBar:2007xeb} & $h_{3d} h_{1\ell}$\\
$B^0 \rightarrow \pi^0  \mu^+ \tau^-$& $4.2 \times 10^{-12}$ & $-$ & $h_{3d} h_{2\ell}$\\
$B^0 \rightarrow \pi^0 \tau^+ e^-$ & $1.2 \times 10^{-9}$ & $-$ & $h_{3d} h_{3\ell}$\\
\hline
$B^0_s \rightarrow \overline{K}^0\,e^+ \mu^-$ & $1.8 \times 10^{-16}$ & $-$ & $h_{3d} h_{1\ell}$ \\
$B^0_s \rightarrow \overline{K}^0 \,\mu^+ \tau^-$& $4.7\times10^{-12}$ & $-$ & $h_{3d} h_{2\ell}$  \\
$B^0_s \rightarrow \overline{K}^0 \, \tau^+ e^-$ & $1.4\times10^{-9}$ & $-$ & $h_{3d} h_{3\ell}$ \\
\hline
\hline
$B_c ^+ \rightarrow D^+ e^+ \mu^-$ &$4.6\times 10^{-17}$ & $-$ & $h_{3d} h_{1\ell}$ \\
$B_c ^+ \rightarrow D^+ \mu^+ \tau^-$ &$1.0\times 10^{-12}$  & $-$ &  $h_{3d} h_{2\ell}$ \\
$B_c ^+ \rightarrow D^+ \tau^+ e^-$ &$3.0\times 10^{-10}$  & $-$ & $h_{3d} h_{3\ell}$ \\
\hline
\hline 
$B^0 \rightarrow \rho^0\, e^+ \mu^-$ & $7.3 \times 10^{-17}$ & $-$  &$h_{3d}h_{1\ell}$\\ 
$B^0 \rightarrow \rho^0\,\mu^+ \tau^-$ & $1.4 \times 10^{-12}$ & $-$ & $h_{3d} h_{2\ell}$ \\
$B^0 \rightarrow \rho^0\,\tau^+ e^-$ & $4.1\times 10^{-10}$ & $-$ & $h_{3d} h_{3\ell}$  \\
\hline
$B^+ \rightarrow \rho^+ \, e^+ \mu^-$& $1.5 \times 10^{-16}$ & $-$ &  $h_{3d} h_{1\ell}$\\
$B^+ \rightarrow \rho^+ \,\mu^+ \tau^-$ & $2.9\times 10^{-12}$  & $-$ & $h_{3d} h_{2\ell}$ \\
$B^+ \rightarrow \rho^+ \,\tau^+ e^-$ & $8.2 \times 10^{-10}$  & $-$ & $h_{3d} h_{3\ell}$ \\
\hline
 $B_s^0 \rightarrow \overline{K}^{0*} e^+ \mu^-$ & $1.5\times
 10^{-16}$& $-$ & $h_{3d} h_{1\ell}$ \\
 $B_s^0 \rightarrow \overline{K}^{0*}\mu^+ \tau^-$ & $2.9\times
 10^{-12}$& $-$ & $h_{3d} h_{2\ell}$ \\
$B_s^0 \rightarrow \bar{K}^{0*}\tau^+ e^-$ & $8.5\times
 10^{-10}$  & $-$ & $h_{3d} h_{3\ell}$\\
\hline

\end{tabular}
\caption{Branching ratios and Yukawa couplings for the CLFV $\bar{b}\rightarrow \bar{d} \, l_m ^+ l_n ^-$ transitions}
\label{b2dllbar}
\end{table}
\begin{table}[h!]
\centering
\small 

\begin{tabular}{|c|c|c|c|}
\hline
& & & \textbf{Dominant} \\
\textbf{Decay Mode} & \textbf{Pred. Limit} & \textbf{Exp. Limit} &\textbf{Yukawa Term} \\
\hline
\hline
$B^+ \rightarrow  K^+ \mu^+ e^-$ & $1.4\times 10^{-11}$ & $6.4 \times 10^{-9}$ \cite{LHCb:2019bix} & $h_{3d} h_{2\ell}$ \\
$B^+ \rightarrow  K^+ \tau^+ \mu^- $& $2.2\times 10^{-9}$  & $5.9 \times 10^{-6}$ \cite{Belle:2022pcr} & $h_{3d} h_{3\ell}$ \\
$B^+ \rightarrow  K^+ e^+ \tau^-$ & $1.2\times 10^{-12}$ & $1.5\times 10^{-5}$\cite{Belle:2022pcr}& $h_{2d} h_{3\ell}$ \\
\hline
$B^0 \rightarrow K^0 \mu^+ e^-$ & $1.5 \times 10^{-11}$  & $3.8 \times 10^{-8}$ 
\cite{BELLE:2019xld} & $h_{3d} h_{2\ell}$ \\
$B^0 \rightarrow K^0 \tau^+ \mu^- $ &$2.3 \times 10^{-9}$ & $-$ & $h_{3d} h_{3\ell}$ \\
$B^0 \rightarrow K^0  e^+ \tau^-$  & $1.2 \times 10^{-12}$  & $-$ & $h_{2d} h_{3\ell}$ \\
\hline
$B^0_s \rightarrow \eta \, \mu^+ e^-$ & $9.5 \times 10^{-12}$ & $-$ & $h_{3d} h_{2\ell}$ \\
$B^0_s \rightarrow \eta \, \tau^+ \mu^-$ & $1.5\times 10^{-9}$  & $-$ & $h_{3d} h_{3\ell}$ \\
$B^0_s \rightarrow \eta \,  e^+ \tau^-$ & $7.8\times 10^{-13}$ & $-$ & $h_{2d} h_{3\ell}$ \\
\hline
$B_c^+ \rightarrow D_s ^+  \mu^+ e^-$ & $2.7\times 10^{-12}$ & $-$ & $h_{3d} h_{2\ell}$\\
$B_c^+ \rightarrow D_s ^+ \tau^+ \mu^-$     & $4.1 \times 10^{-10}$  &
$-$& $h_{3d} h_{3\ell}$\\
$B_c^+ \rightarrow D_s ^+  e^+ \tau^-$     & $2.1 \times 10^{-13}$  & 
$-$& $h_{2d} h_{3\ell}$ \\
\hline
\hline
$B^0 \rightarrow K^{0*} \mu^+ e^-$ & $ 5.3 \times 10^{-12}$ & $ 5.7 \times10^{-9}$ 
\cite{LHCb:2022lrd} & $h_{3d} h_{2\ell}$ \\ 
$B^0 \rightarrow K^{0*}\tau^+ \mu^-$ & $  6.2 \times 10^{-10}$ & $ 1.0\times10^{-5}$ 
\cite{LHCb:2022wrs} & $h_{3d} h_{3\ell}$ \\
$B^0 \rightarrow K^{0*} e^+ \tau^-$ & $3.3\times 10^{-13}$ & $5.9 \times 10^{-6} $ \cite{LHCb:2025eyf} & $h_{2d} h_{3\ell}$ \\
\hline
 $B_s \rightarrow \phi \, \mu^+ e^-$  & $6.0\times10^{-12}$ & $ 1.6 \times 10^{-8}$ \cite{LHCb:2022lrd} & $h_{3d} h_{2\ell}$ \\
$B_s \rightarrow \phi \, \tau^+ \mu^-$ & $7.0\times 10^{-10}$ &  $ 1.0 \times 10^{-5}$ \cite{LHCb:2024wve} & $h_{3d} h_{3\ell}$ \\
$B_s \rightarrow \phi \, e^+ \tau^-$  & $3.6\times 10^{-13}$ & $-$ & $h_{2d} h_{3\ell}$ \\
\hline
\end{tabular}
\caption{Branching Ratios and Yukawa Couplings for the CLFV $\bar{b}\rightarrow \bar{s} \, l_m ^+ l_n ^-$ transitions}
\label{b2sllbar}
\end{table}
\subsection{CLFV of up-type quarks}
Only in recent times, flavour-changing neutral interactions in charm meson decays are
studied in detail. The present experimental bounds on the branching ratios of some of these decays are $1.3 \times 10^{-8}$ for $D^0 \to \mu^- e^+$ \cite{LHCb:2015pce} and $8.0 \times 10^{-7}$ for $D^0 \to \pi^0 \mu^- e^+$ \cite{BaBar:2020faa}.
We have computed the rates for these branching ratios and other related decays such as $D^+ \rightarrow \pi^+ \mu^- e^+$ and $D_s^+ \rightarrow K^+ \mu^- e^+$ in the present $A_4$ model, with the constraint on the masses given in eq \ref{Phi23masslimit}. In each case, we obtained an upper bound on the branching ratio of the order of $10^{-18}$. The CLFV in the charm sector is negligibly small. 

The flavour-violating decays of the top quark can be substantial in this model because the amplitudes will be proportional to the large top quark Yukawa coupling. 
From the couplings listed in table~\ref{yucup}, we pick the CLFV decays of top
quark, whose amplitudes are proportional to $(h_t\, h_\tau)$, which is the product of the largest Yukawa couplings in the quark sector and the lepton sector. 
These decays are $t \to c \, \tau^+  e^-$ and $t \to u \, \tau^+ \mu^-$.
LHC experiments have searched for CLFV decays of top quarks \cite{Watson:2025ajf}. CMS experiment has set a very stringent bound on the scalar-mediated CLFV branching ratio of the top quark $BR (t \to c\, \mu \, \tau) < 8 \times 10^{-7}$ \cite{CMS:2025xnf}. The bounds on CLFV branching ratios of top quark to a final state with $e + \tau$ leptons, mediated by scalars, are also very stringent with $BR(t \to u\, \tau \mu) < 4 \times 10^{-8}$ \cite{CMS:2025xnf}. We calculate the upper bounds on these two decays to be 
$1.3 \times 10^{-9}$. These bounds are one to two orders of magnitude smaller than the current experimental bounds. Observing these CLFV top decays in future at LHC can be a pointer to the validity of this model.

\subsection{CLFV of leptons}\label{radiative}

\begin{figure}[h]
    \begin{minipage}[c]{0.5\textwidth}
    \centering
    \includegraphics[scale=.5]{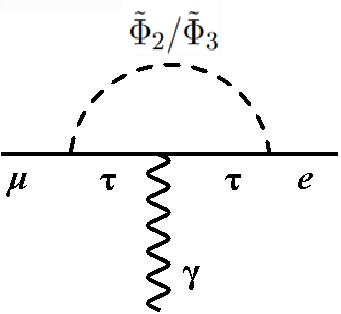}
    \end{minipage}
    \begin{minipage}[c]{0.5\textwidth}
    \includegraphics[scale=.5]{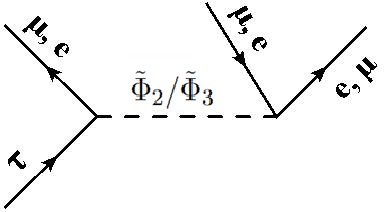}
    \end{minipage}
\caption{Feynman diagram for the radiative CLFV decay
    $\mu \to e \, \gamma$ on the left panel and $\tau$-lepton decay on the right panel}\label{radclfv}
\end{figure}

The Feynman diagram for the radiative CLFV decay $\mu \to e \,\gamma$, mediated by $\tilde{\Phi}_2,
\tilde{\Phi}_3$, is shown in the left panel of
fig.~\ref{radclfv}.
The amplitude for this diagram is  
\be
\mathcal{A} (\mu \to e \, \gamma) \propto (g_2^{e \tau})(g_2^{\mu \tau})^*\left[\frac{1}{m_{\Re\tilde{\Phi}_2}^2}+\frac{1}{m_{\Im\tilde{\Phi}_2}^2}\right]+(g_3^{e \tau})(g_3^{\mu \tau})^*\left[\frac{1}{m_{\Re\tilde{\Phi}_3}^2}+\frac{1}{m_{\Im\tilde{\Phi}_3}^2}\right].
\ee
From table~\ref{clfvcff}, we see that the product of the coefficients, both for $\tilde{\Phi}_2$ and $\tilde{\Phi}_3$ exchange, in the above amplitude vanishes. The same consideration applies to $\tau \to \mu \, \gamma$ and $\tau \to e \, \gamma$. This vanishing amplitude for the neutral scalar-mediated radiative CLFV decays is, once again, a consequence of the $A_4$ symmetry of the model.

The Feynman diagram for $(g-2)_\mu$ will be very similar to that in fig.~\ref{radclfv}, except that both external leptons will be muons. For such a diagram, the amplitude will be non-zero because it involves a product of two non-zero coefficients. However, substituting the bound in eq.~\ref{Phi23masslimit}, we find that the contribution of this diagram to $(g-2)_\mu$ will be about $10^{-13}$, much smaller than the present experimental uncertainty~\cite{Muong-2:2023cdq}.

This model also predicts two very interesting and unique CLFV decays of $\tau$ lepton, $\tau^- \to \mu^- \mu^- e^+$
and $\tau^- \to e^- e^- \mu^+$, shown in the right panel of fig.~\ref{radclfv}. The calculated upper limit on the branching ratio for both these modes is of the order of $10^{-14}$, which is six orders of magnitude smaller than the present experimental bound of $1.5 \times 10^{-8}$ \cite{Hayasaka:2010np}.

\section{Conclusion}\label{conclusion}
The interrelations between CLFV decays involving the same flavours of charged leptons depend on the mechanism of neutrino mass generation. If this mechanism consists only of seesaw and no other BSM physics, then it is shown \cite{Awasthi:2024nvi} that the branching ratios of the radiative CLFV decays are always larger than those of the corresponding meson CLFV decays. It is very interesting to study models of neutrino mass generation which are in contrast to the above statement. Here we explored one such model, which is based on $A_4$ flavour symmetry.

In this work, we consider the most general CP-conserving scalar potential of the model and obtain the neutral scalar mass eigenstates. We find that the model contains two neutral scalar mass eigenstates with purely flavour-changing interactions to charged fermions.
The $A_4$ symmetry of the model constrains these interactions in such a way that they mediate meson CLFV transitions but not radiative CLFV transitions. Thus, this model is an example of a rare case where the branching ratios of meson CLFV decays are larger than those of the corresponding radiative CLFV decays.

We derive a bound on the masses of the flavour-violating neutral scalars from the branching ratio of the decay $K_L \to \mu^+ e^-$. Utilizing this bound, we predict the upper limits on the purely leptonic and semileptonic CLFV decay modes of various mesons.
An important consequence of the $A_4$ symmetry of the model is the distinctive charge selection rule of meson CLFV decays. For example, if $M^0 \rightarrow \ell_m^+ \ell_n^-$ is allowed then $M^0 \rightarrow \ell_m^- \ell_n^+$ is forbidden. We studied the purely leptonic CLFV decays of this model in an earlier work~\cite{Korrapati:2020rao}, where a much simpler form of the scalar potential was considered. However, the pattern of CLFV decays remains the same whether the simpler or the full form of the scalar potential is considered. The upper bounds on the CLFV branching ratios for the two forms of the potential were found to be very similar.

The predictions of the model for the following five CLFV decays:
\begin{itemize}
    \item $K_L \to \mu^+ e^-$ 
    \item $B_s^0 \to \mu^+ e^-$
    \item $B^+ \to K^+ \mu^+ e^-$
    \item $t \to c \, \tau^+  e^-$ 
    \item $t \to u \, \tau^+ \mu^-$
\end{itemize}
are one to two orders of magnitude below the present experimental limits and are likely to be accessible at the next generation of experiments.

\section*{Acknowledgements} 
Jai More thanks the
Department of Science and Technology (DST), Government of India, for the financial support through the grant
no. DST/WISE-PDF/PM-91/2023(G). Akhila Kumar Pradhan, Purushottam Sahu and S. Uma Sankar thank the Ministry of Education, Government of India, for financial support through Institute of Eminence funding to I.I.T. Bombay.
\appendix
\section{Neutral scalar mass matrix}\label{NSmassmatrix}

Applying the $A_4$ algebra, we can write the potential in terms of the Higgs fields ($\phi_0, \phi_1, \phi_2, \phi_3$) as follows:
\be
V(\phi_i)\es \mu_1^2 \left(\psq{1}+\psq{2}+\psq{3}\right)+\Lambda_1\left(\phi_1^4+\phi_2^4+\phi_3^4\right)+ \Lambda_2\left(\psq{1}\ \psq{2}+\psq{2} \ \psq{3}+\psq{3}\ \psq{1}\right)\nn
\ps
\Lambda_3\left((\pdagp{1}{2})^2+(\pdagp{2}{3})^2+(\pdagp{3}{1})^2\right)+\Lambda_3^*\left((\pdagp{2}{1})^2+(\pdagp{3}{2})^2+(\pdagp{1}{3})^2\right)
\nn
\ps
\Lambda_4\left(\pdagp{2}{1}\pdagp{1}{2}+\pdagp{3}{2}\pdagp{2}{3}+\pdagp{1}{3}\pdagp{3}{1}\right)+\Lambda_4^*\left(\pdagp{1}{2}\pdagp{2}{1}+\pdagp{2}{3}\pdagp{3}{2}+\pdagp{3}{1}\pdagp{1}{3}\right),\nn
\ee
here $\phi_\alpha^2=\phi_\alpha^\dagger\phi_\alpha$, $~~\phi_\alpha^4=(\phi_\alpha^\dagger\phi_\alpha)^2$,~ $\alpha \equiv \{0,1,2,3\}$
\be
\Lambda_1\es\lambda_1+\lambda_2,\nn
\Lambda_2\es 2\lambda_1 -\lambda_2,\nn
\Lambda_3\es\lambda_3+\lambda_4+i\lambda_5,\nn
\Lambda_4\es\lambda_3-\lambda_4+i\lambda_5,\\
V(\phi_0)\es \mu_2^2 \phi^2_0 +\lambda \phi^4_0,\\
V(\chi_i)\es \mu_3^2 (\chi_1^2+\chi_2^2+\chi_3^2)+ 6~\eta_1 \chi_1 \chi_2 \chi_3+\eta_2 (\chi_1^2+\chi_2^2+\chi_3^2)^2\nn
\ps \eta_3 (\chi_1^2+\omega \chi_2^2+ \omega^2 \chi_3^2)(\chi_1^2+\omega^2 \chi_2^2+ \omega \chi_3^2)+ 4\eta_4 (\chi_2^2 \chi_3^2+\chi_3^2 \chi_1^2+\chi_1^2 \chi_2^2),\\
V(\phi_i,\phi_0)\es \rho_1 \phi^2_0\left(\psq{1}+\psq{2}+\psq{3}\right)+\rho_2\left(\pdagp{1}{0}\pdagp{0}{1}+\pdagp{2}{0}\pdagp{0}{2}+\pdagp{3}{0}\pdagp{0}{3}\right)\nn
\ps \rho_3 \left(\pdagp{1}{0}\pdagp{1}{0}+\pdagp{2}{0}\pdagp{2}{0}+\pdagp{3}{0}\pdagp{3}{0}\right)+\rho_3^* \left(\pdagp{0}{1}\pdagp{0}{1}+\pdagp{0}{2}\pdagp{0}{2}+\pdagp{0}{3}\pdagp{0}{3}\right),\nn
V(\phi_i, \chi_i) \es 
\delta_3 (\pdagp{1}{1}+\pdagp{2}{2}+\pdagp{3}{3})(\chi_1^2+\chi_2^2+\chi_3^2)\nn
\ps 2\delta_5 \left((\phi_2^\dagger \phi_3+\phi_3^\dagger \phi_2)\chi_2\chi_3+(\phi_3^\dagger \phi_1+\phi_1^\dagger \phi_3)\chi_1\chi_3+(\phi_1^\dagger \phi_2+\phi_2^\dagger \phi_1)\chi_1\chi_2\right),\\
V(\phi_0,\chi_i)\es \sigma \phi_0^\dagger \phi_0 (\chi_1^2+\chi_2^2+\chi_3^2).
\ee

Let us now discuss some key observations regarding the mixing between $\Re\Phi_{2,3}$ and $\chi_i$ arising due to $\mathcal{M}^2_{\Re \Phi \chi}$ in eq.~\ref{Mphichi}. Independent of the value of $\delta_5$, it can be shown that $ \langle \chi_2\vert\mathcal{M}^2_{\Re \Phi_\alpha \chi} \vert \Re \Phi_{2,3}\rangle = 0$.
In the limit $\delta_5=0$,  $\mathcal{M}^2_{\chi \chi}$ is diagonal and 
$\langle \chi_{1,3}\vert\mathcal{M}^2_{\Re \Phi_\alpha \chi} \vert \Re\Phi_{2,3}\rangle = 0.$  
Hence, there is no admixture of $\chi_i$ with $\Re \Phi_{2,3}$  and their masses remain degenerate. Any admixture of $\chi_i$ with $\Re\Phi_{2,3}$ has to be proportional to $\delta_5$. This admixture can break the degeneracy of the masses of $\Re \Phi_{2,3}$. The extremely small value the CP violation in $K^0-\bar{K}^0$ mixing imposes the constraint \cite{Korrapati:2020rao}.
\be\label{massdiff}
\frac{m_{\Re \Phi_2}-m_{\Re \Phi_3}}{m_{\Re\Phi_2}} \leq 10^{-3}.
\ee
Since the non-zero value of $\delta_5$ is responsible for the small mass difference in the numerator of eq.~\ref{massdiff}, its magnitude has to be small. 
Such small values of $\delta_5$ do not have any appreciable impact on the phenomenology of tree-level CLFV decays.
\section{Estimation of neutrino mass}\label{neutrinomass}
The heavy Majorana neutrino mass eigenvalues are $M-M', M, M+M'$, leading to light neutrino masses for normal hierarchy (NH) 
\be
m_1 = \frac{m_D^2}{M+M'},~~~~~m_2 = \frac{m_D^2}{M},~~~~m_3 = \frac{m_D^2}{M-M'}
\ee
where $m_D$
is the common Dirac mass. The cosmological limit on the sum of neutrino masses,
    \be
\label{sumofneutrino}    m_1+m_2+m_3 <0.11 ~\text{eV}
 \ee
 imposes the constraint
 \be 
 \frac{m_D^2}{M}\left(\frac{3-\Delta^2}{1-\Delta^2} \right)<0.11, ~~~~ \Delta=\frac{M'}{M}
\ee
For NH, the ratio of mass-squared splittings,  
\be \label{NHsplit}
\frac{\Delta_{21}}{\Delta_{31}} \es \frac{m_2^2-m_1^2}{m_3^2-m_1^2}\nn
\ee
reduces to 
\be
\frac1{4}(2+\Delta)(1-\Delta)^2  = 0.03
\ee
which matches the observed value $\Delta_{21}/\Delta_{31} \simeq 0.03$ for $\Delta \simeq 0.9$. This, with the cosmological bound in eq.~\ref{sumofneutrino}, gives 
$m_D \simeq 0.1$ ~MeV and $M \simeq 1$~ TeV.
\begin{figure}[h]    \centering
\includegraphics[scale=0.4]{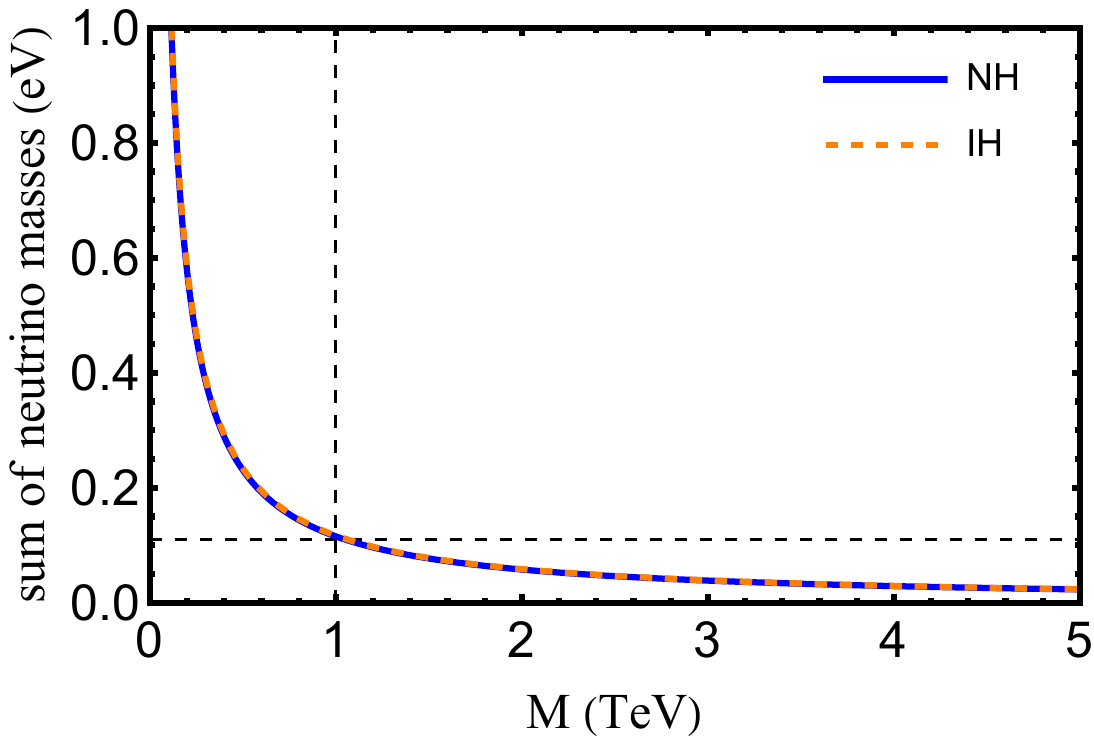}
\label{massplot}
\caption{This plot gives a study of Majorana mass $M$ using the cosmological constraint on neutrino masses. The horizontal dashed line denotes the cosmological upper limit, while the vertical dashed line shows the corresponding bound on $M$.}
\end{figure}

For the inverted hierarchy (IH), the masses are

\be
m_1 = \frac{m_D^2}{M},~~~~m_2 = \frac{m_D^2}{M-M'},~~~~~m_3 = \frac{m_D^2}{M+M'},
\ee
and the ratio is
\be\label{IHsplit}
\frac{\Delta_{23}}{\Delta_{21}} 
\es \frac{m_2^2-m_3^2}{m_2^2-m_1^2}
\ee
becomes
\be\label{constraint}
\frac{4}{(2-\Delta)(1+\Delta)^2 } \approx 30
\ee
The cosmological constraint in eq.~\ref{sumofneutrino} with the constraint in eq.~\ref{constraint}, yields $\Delta \simeq -1.99, m_D=0.6$ MeV and $M=1$ TeV.
\bibliographystyle{utcaps_mod}
\bibliography{reference.bib}

\providecommand{\href}[2]{#2}\begingroup\raggedright\begin{thebibliography}{10}

\bibitem{MEGII:2023ltw}
{\normalfont \bfseries MEG II}, K.~Afanaciev {\em et al.}, ``{\em {A search for $\mu^+ \rightarrow e^+ \gamma$ with the first dataset of the MEG~II experiment}},'' \href{http://dx.doi.org/10.1140/epjc/s10052-024-12416-2}{Eur. Phys. J. C {\normalfont \bfseries 84} (2024) no.~3, 216}, \href{http://arxiv.org/abs/2310.12614}{{\normalfont \ttfamily arXiv:2310.12614}}. [Erratum: Eur.Phys.J.C 84, 1042 (2024)].

\bibitem{COMET:2018auw}
{\normalfont \bfseries COMET}, R.~Abramishvili {\em et al.}, ``{\em {COMET Phase-I Technical Design Report}},'' \href{http://dx.doi.org/10.1093/ptep/ptz125}{PTEP {\normalfont \bfseries 2020} (2020) no.~3, 033C01}, \href{http://arxiv.org/abs/1812.09018}{{\normalfont \ttfamily arXiv:1812.09018}}.

\bibitem{Mu3e:2020gyw}
{\normalfont \bfseries Mu3e}, K.~Arndt {\em et al.}, ``{\em {Technical design of the phase I Mu3e experiment}},'' \href{http://dx.doi.org/10.1016/j.nima.2021.165679}{Nucl. Instrum. Meth. A {\normalfont \bfseries 1014} (2021)  165679}, \href{http://arxiv.org/abs/2009.11690}{{\normalfont \ttfamily arXiv:2009.11690}}.

\bibitem{King:2003jb}
S.~F. King, ``{\em {Neutrino mass models}},'' \href{http://dx.doi.org/10.1088/0034-4885/67/2/R01}{Rept. Prog. Phys. {\normalfont \bfseries 67} (2004)  107--158}, \href{http://arxiv.org/abs/hep-ph/0310204}{{\normalfont \ttfamily arXiv:hep-ph/0310204}}.

\bibitem{Awasthi:2024nvi}
P.~C. Awasthi, J.~More, A.~K. Pradhan, K.~Rao, P.~Sahu, and S.~U. Sankar, ``{\em {Charged Lepton Flavour Violating meson decays in seesaw models}},'' \href{http://dx.doi.org/10.1007/JHEP03(2025)183}{JHEP {\normalfont \bfseries 03} (2025)  183}, \href{http://arxiv.org/abs/2410.10490}{{\normalfont \ttfamily arXiv:2410.10490}}.

\bibitem{He:2006dk}
X.-G. He, Y.-Y. Keum, and R.~R. Volkas, ``{\em {A(4) flavor symmetry breaking scheme for understanding quark and neutrino mixing angles}},'' \href{http://dx.doi.org/10.1088/1126-6708/2006/04/039}{JHEP {\normalfont \bfseries 04} (2006)  039}, \href{http://arxiv.org/abs/hep-ph/0601001}{{\normalfont \ttfamily arXiv:hep-ph/0601001}}.

\bibitem{DayaBay:2012fng}
{\normalfont \bfseries Daya Bay}, F.~P. An {\em et al.}, ``{\em {Observation of electron-antineutrino disappearance at Daya Bay}},'' \href{http://dx.doi.org/10.1103/PhysRevLett.108.171803}{Phys. Rev. Lett. {\normalfont \bfseries 108} (2012)  171803}, \href{http://arxiv.org/abs/1203.1669}{{\normalfont \ttfamily arXiv:1203.1669}}.

\bibitem{RENO:2012mkc}
{\normalfont \bfseries RENO}, J.~K. Ahn {\em et al.}, ``{\em {Observation of Reactor Electron Antineutrino Disappearance in the RENO Experiment}},'' \href{http://dx.doi.org/10.1103/PhysRevLett.108.191802}{Phys. Rev. Lett. {\normalfont \bfseries 108} (2012)  191802}, \href{http://arxiv.org/abs/1204.0626}{{\normalfont \ttfamily arXiv:1204.0626}}.

\bibitem{Esteban:2024eli}
I.~Esteban, M.~C. Gonzalez-Garcia, M.~Maltoni, I.~Martinez-Soler, J.~P. Pinheiro, and T.~Schwetz, ``{\em {NuFit-6.0: updated global analysis of three-flavor neutrino oscillations}},'' \href{http://dx.doi.org/10.1007/JHEP12(2024)216}{JHEP {\normalfont \bfseries 12} (2024)  216}, \href{http://arxiv.org/abs/2410.05380}{{\normalfont \ttfamily arXiv:2410.05380}}.

\bibitem{Dev:2015dha}
A.~Dev, P.~Ramadevi, and S.~U. Sankar, ``{\em {Non-zero {\ensuremath{\theta}}$_{13}$ and {\ensuremath{\delta}}$_{CP}$ in a neutrino mass model with A$_{4}$ symmetry}},'' \href{http://dx.doi.org/10.1007/JHEP11(2015)034}{JHEP {\normalfont \bfseries 11} (2015)  034}, \href{http://arxiv.org/abs/1504.04034}{{\normalfont \ttfamily arXiv:1504.04034}}.

\bibitem{Korrapati:2020rao}
R.~Korrapati, J.~More, U.~Rahaman, and S.~U. Sankar, ``{\em {Signatures of $A_4$ symmetry in the charged lepton flavour violating decays in a neutrino mass model}},'' \href{http://dx.doi.org/10.1140/epjc/s10052-021-09171-z}{Eur. Phys. J. C {\normalfont \bfseries 81} (2021) no.~5, 382}, \href{http://arxiv.org/abs/2009.00865}{{\normalfont \ttfamily arXiv:2009.00865}}.

\bibitem{Grimus:2011fk}
W.~Grimus and P.~O. Ludl, ``{\em {Finite flavour groups of fermions}},'' \href{http://dx.doi.org/10.1088/1751-8113/45/23/233001}{J. Phys. A {\normalfont \bfseries 45} (2012)  233001}, \href{http://arxiv.org/abs/1110.6376}{{\normalfont \ttfamily arXiv:1110.6376}}.

\bibitem{Planck:2018vyg}
{\normalfont \bfseries Planck}, N.~Aghanim {\em et al.}, ``{\em {Planck 2018 results. VI. Cosmological parameters}},'' \href{http://dx.doi.org/10.1051/0004-6361/201833910}{Astron. Astrophys. {\normalfont \bfseries 641} (2020)  A6}, \href{http://arxiv.org/abs/1807.06209}{{\normalfont \ttfamily arXiv:1807.06209}}. [Erratum: Astron.Astrophys. 652, C4 (2021)].

\bibitem{Minkowski:1977sc}
P.~Minkowski, ``{\em {$\mu \to e\gamma$ at a Rate of One Out of $10^{9}$ Muon Decays?}},''
\href{http://dx.doi.org/10.1016/0370-2693(77)90435-X}{Phys. Lett. {\normalfont \bfseries 67B} (1977)  421--428}.

\bibitem{Dinh:2012bp}
D.~N. Dinh, A.~Ibarra, E.~Molinaro, and S.~T. Petcov, ``{\em {The $\mu - e$ Conversion in Nuclei, $\mu \to e \gamma, \mu \to 3e$ Decays and TeV Scale See-Saw Scenarios of Neutrino Mass Generation}},'' \href{http://dx.doi.org/10.1007/JHEP08(2012)125}{JHEP {\normalfont \bfseries 08} (2012)  125}, \href{http://arxiv.org/abs/1205.4671}{{\normalfont \ttfamily arXiv:1205.4671}}. [Erratum: JHEP 09, 023 (2013)].

\bibitem{BNL:1998apv}
{\normalfont \bfseries BNL}, D.~Ambrose {\em et al.}, ``{\em {New limit on muon and electron lepton number violation from $K^0_L \rightarrow \mu^\pm e^\mp$ decay}},'' \href{http://dx.doi.org/10.1103/PhysRevLett.81.5734}{Phys. Rev. Lett. {\normalfont \bfseries 81} (1998)  5734--5737}, \href{http://arxiv.org/abs/hep-ex/9811038}{{\normalfont \ttfamily arXiv:hep-ex/9811038}}.

\bibitem{ATLAS:2024vxm}
{\normalfont \bfseries ATLAS}, G.~Aad {\em et al.}, ``{\em {Search for heavy neutral Higgs bosons decaying into a top quark pair in 140 fb$^{-1}$ of proton-proton collision data at $ \sqrt{s} $ = 13 TeV with the ATLAS detector}},'' \href{http://dx.doi.org/10.1007/JHEP08(2024)013}{JHEP {\normalfont \bfseries 08} (2024)  013}, \href{http://arxiv.org/abs/2404.18986}{{\normalfont \ttfamily arXiv:2404.18986}}.

\bibitem{CMS:2019pex}
{\normalfont \bfseries CMS}, A.~M. Sirunyan {\em et al.}, ``{\em {Search for lepton flavour violating decays of a neutral heavy Higgs boson to $\mu\tau$ and e$\tau$ in proton-proton collisions at $\sqrt{s}=$ 13 TeV}},'' \href{http://dx.doi.org/10.1007/JHEP03(2020)103}{JHEP {\normalfont \bfseries 03} (2020)  103}, \href{http://arxiv.org/abs/1911.10267}{{\normalfont \ttfamily arXiv:1911.10267}}.

\bibitem{LHCb:2017hag}
{\normalfont \bfseries LHCb}, R.~Aaij {\em et al.}, ``{\em {Search for the lepton-flavour violating decays $B_{(s)}^{0} \to e^\pm \mu^\mp $}},'' \href{http://dx.doi.org/10.1007/JHEP03(2018)078}{JHEP {\normalfont \bfseries 03} (2018)  078}, \href{http://arxiv.org/abs/1710.04111}{{\normalfont \ttfamily arXiv:1710.04111}}.

\bibitem{LHCb:2019ujz}
{\normalfont \bfseries LHCb}, R.~Aaij {\em et al.}, ``{\em {Search for the lepton-flavour-violating decays $B^{0}_{s}\to\tau^{\pm}\mu^{\mp}$ and $B^{0}\to\tau^{\pm}\mu^{\mp}$}},'' \href{http://dx.doi.org/10.1103/PhysRevLett.123.211801}{Phys. Rev. Lett. {\normalfont \bfseries 123} (2019) no.~21, 211801}, \href{http://arxiv.org/abs/1905.06614}{{\normalfont \ttfamily arXiv:1905.06614}}.

\bibitem{Belle:2021rod}
{\normalfont \bfseries Belle}, H.~Atmacan {\em et al.}, ``{\em {Search for $B^{0} \to \tau^\pm \ell^\mp$ ($\ell=e,\mu$) with a hadronic tagging method at Belle}},'' \href{http://dx.doi.org/10.1103/PhysRevD.104.L091105}{Phys. Rev. D {\normalfont \bfseries 104} (2021) no.~9, L091105}, \href{http://arxiv.org/abs/2108.11649}{{\normalfont \ttfamily arXiv:2108.11649}}.

\bibitem{Belle:2023jwr}
{\normalfont \bfseries Belle}, L.~Nayak {\em et al.}, ``{\em {Search for $B^0_s \rightarrow \ell^{\mp} \tau^{\pm}$ with the Semi-leptonic Tagging Method at Belle}},'' \href{http://dx.doi.org/10.1007/JHEP08(2023)178}{JHEP {\normalfont \bfseries 08} (2023)  178}, \href{http://arxiv.org/abs/2301.10989}{{\normalfont \ttfamily arXiv:2301.10989}}.

\bibitem{Carrasco:2016kpy}
N.~Carrasco, P.~Lami, V.~Lubicz, L.~Riggio, S.~Simula, and C.~Tarantino, ``{\em {$K \to \pi$ semileptonic form factors with $N_f=2+1+1$ twisted mass fermions}},'' \href{http://dx.doi.org/10.1103/PhysRevD.93.114512}{Phys. Rev. D {\normalfont \bfseries 93} (2016) no.~11, 114512}, \href{http://arxiv.org/abs/1602.04113}{{\normalfont \ttfamily arXiv:1602.04113}}.

\bibitem{Becirevic:2024vwy}
D.~Be{\v{c}}irevi{\'c}, F.~Jaffredo, J.~P. Pinheiro, and O.~Sumensari, ``{\em {Lepton flavor violation in exclusive b{\textrightarrow}d{\ensuremath{\ell}}i{\ensuremath{\ell}}j and b{\textrightarrow}s{\ensuremath{\ell}}i{\ensuremath{\ell}}j decay modes}},'' \href{http://dx.doi.org/10.1103/PhysRevD.110.075004}{Phys. Rev. D {\normalfont \bfseries 110} (2024) no.~7, 075004}, \href{http://arxiv.org/abs/2407.19060}{{\normalfont \ttfamily arXiv:2407.19060}}.

\bibitem{Sher:2005sp}
A.~Sher {\em et al.}, ``{\em {An Improved upper limit on the decay $K^+ \rightarrow \pi^+ \mu^+ e^-$}},'' \href{http://dx.doi.org/10.1103/PhysRevD.72.012005}{Phys. Rev. D {\normalfont \bfseries 72} (2005)  012005}, \href{http://arxiv.org/abs/hep-ex/0502020}{{\normalfont \ttfamily arXiv:hep-ex/0502020}}.

\bibitem{Weir:1989sq}
A.~J. Weir {\em et al.}, ``{\em {Upper Limits on $D^\pm$ and $B^\pm$ Decays to Two Leptons Plus $\pi^\pm$ or $K^\pm$}},'' \href{http://dx.doi.org/10.1103/PhysRevD.41.1384}{Phys. Rev. D {\normalfont \bfseries 41} (1990)  1384}.

\bibitem{BaBar:2012azg}
{\normalfont \bfseries BaBar}, J.~P. Lees {\em et al.}, ``{\em {A search for the decay modes $B^{+-} \to h^{+-} \tau^{+-}l$}},'' \href{http://dx.doi.org/10.1103/PhysRevD.86.012004}{Phys. Rev. D {\normalfont \bfseries 86} (2012)  012004}, \href{http://arxiv.org/abs/1204.2852}{{\normalfont \ttfamily arXiv:1204.2852}}.

\bibitem{BaBar:2007xeb}
{\normalfont \bfseries BaBar}, B.~Aubert {\em et al.}, ``{\em {Search for the rare decay $B \to \pi l^+ l^-$}},'' \href{http://dx.doi.org/10.1103/PhysRevLett.99.051801}{Phys. Rev. Lett. {\normalfont \bfseries 99} (2007)  051801}, \href{http://arxiv.org/abs/hep-ex/0703018}{{\normalfont \ttfamily arXiv:hep-ex/0703018}}.

\bibitem{LHCb:2019bix}
{\normalfont \bfseries LHCb}, R.~Aaij {\em et al.}, ``{\em {Search for Lepton-Flavor Violating Decays $B^+ \to K^+ {\mu}^{\pm} e^{\mp}$}},'' \href{http://dx.doi.org/10.1103/PhysRevLett.123.241802}{Phys. Rev. Lett. {\normalfont \bfseries 123} (2019) no.~24, 241802}, \href{http://arxiv.org/abs/1909.01010}{{\normalfont \ttfamily arXiv:1909.01010}}.

\bibitem{Belle:2022pcr}
{\normalfont \bfseries Belle}, S.~Watanuki {\em et al.}, ``{\em {Search for the Lepton Flavor Violating Decays $ B^+ \rightarrow K^+ \tau^\pm\ell^\mp (\ell =e, \mu) $ at Belle}},'' \href{http://dx.doi.org/10.1103/PhysRevLett.130.261802}{Phys. Rev. Lett. {\normalfont \bfseries 130} (2023) no.~26, 261802}, \href{http://arxiv.org/abs/2212.04128}{{\normalfont \ttfamily arXiv:2212.04128}}.

\bibitem{BELLE:2019xld}
{\normalfont \bfseries BELLE}, S.~Choudhury {\em et al.}, ``{\em {Test of lepton flavor universality and search for lepton flavor violation in $B \rightarrow K\ell \ell$ decays}},'' \href{http://dx.doi.org/10.1007/JHEP03(2021)105}{JHEP {\normalfont \bfseries 03} (2021)  105}, \href{http://arxiv.org/abs/1908.01848}{{\normalfont \ttfamily arXiv:1908.01848}}.

\bibitem{LHCb:2022lrd}
{\normalfont \bfseries LHCb}, R.~Aaij {\em et al.}, ``{\em {Search for the lepton-flavour violating decays $B^{0} \rightarrow K^{*0} \mu^\pm e^\mp$ and $B_s^0 \rightarrow \phi \mu^\pm e^\mp$}},'' \href{http://dx.doi.org/10.1007/JHEP06(2023)073}{JHEP {\normalfont \bfseries 06} (2023)  073}, \href{http://arxiv.org/abs/2207.04005}{{\normalfont \ttfamily arXiv:2207.04005}}.

\bibitem{LHCb:2022wrs}
{\normalfont \bfseries LHCb}, R.~Aaij {\em et al.}, ``{\em {Search for the lepton-flavour violating decays $B^0 \to K^{*0} \tau^\pm \mu^\mp$}},'' \href{http://dx.doi.org/10.1007/JHEP06(2023)143}{JHEP {\normalfont \bfseries 06} (2023)  143}, \href{http://arxiv.org/abs/2209.09846}{{\normalfont \ttfamily arXiv:2209.09846}}.

\bibitem{LHCb:2025eyf}
{\normalfont \bfseries LHCb}, R.~Aaij {\em et al.}, ``{\em {Search for the lepton-flavour-violating decays $B^0 \to K^{*0} \tau^\pm e^\mp$}},'' \href{http://arxiv.org/abs/2506.15347}{{\normalfont \ttfamily arXiv:2506.15347}}.

\bibitem{LHCb:2024wve}
{\normalfont \bfseries LHCb}, R.~Aaij {\em et al.}, ``{\em {Search for the lepton-flavor violating decay $B_s^0\rightarrow \phi \mu^\pm\tau^\mp$}},'' \href{http://dx.doi.org/10.1103/PhysRevD.110.072014}{Phys. Rev. D {\normalfont \bfseries 110} (2024) no.~7, 072014}, \href{http://arxiv.org/abs/2405.13103}{{\normalfont \ttfamily arXiv:2405.13103}}.

\bibitem{LHCb:2015pce}
{\normalfont \bfseries LHCb}, R.~Aaij {\em et al.}, ``{\em {Search for the lepton-flavour violating decay $D^0 \to e^\pm \mu^\mp$}},'' \href{http://dx.doi.org/10.1016/j.physletb.2016.01.029}{Phys. Lett. B {\normalfont \bfseries 754} (2016)  167--175}, \href{http://arxiv.org/abs/1512.00322}{{\normalfont \ttfamily arXiv:1512.00322}}.

\bibitem{BaBar:2020faa}
{\normalfont \bfseries BaBar}, J.~P. Lees {\em et al.}, ``{\em {Search for lepton-flavor-violating decays $D^{0}\rightarrow X^{0}e^{\pm}\mu^{\mp}$}},'' \href{http://dx.doi.org/10.1103/PhysRevD.101.112003}{Phys. Rev. D {\normalfont \bfseries 101} (2020) no.~11, 112003}, \href{http://arxiv.org/abs/2004.09457}{{\normalfont \ttfamily arXiv:2004.09457}}.

\bibitem{Watson:2025ajf}
{\normalfont \bfseries ATLAS, CMS}, M.~Watson, ``{\em {Probes of flavour symmetry and violation with top quarks in ATLAS and CMS}},'' in {\em {17th International Workshop on Top Quark Physics}}.
\newblock 1, 2025.
\newblock \href{http://arxiv.org/abs/2501.14498}{{\normalfont \ttfamily arXiv:2501.14498}}.

\bibitem{CMS:2025xnf}
{\normalfont \bfseries CMS}, A.~Hayrapetyan {\em et al.}, ``{\em {Search for charged-lepton flavour violation in top quark interactions with an up-type quark, a muon, and a $\tau$ lepton in proton-proton collisions at $\sqrt{s}$ = 13 TeV}},'' \href{http://arxiv.org/abs/2504.08532}{{\normalfont \ttfamily arXiv:2504.08532}}.

\bibitem{Muong-2:2023cdq}
{\normalfont \bfseries Muon g-2}, D.~P. Aguillard {\em et al.}, ``{\em {Measurement of the Positive Muon Anomalous Magnetic Moment to 0.20~ppm}},'' \href{http://dx.doi.org/10.1103/PhysRevLett.131.161802}{Phys. Rev. Lett. {\normalfont \bfseries 131} (2023) no.~16, 161802}, \href{http://arxiv.org/abs/2308.06230}{{\normalfont \ttfamily arXiv:2308.06230}}.

\bibitem{Hayasaka:2010np}
K.~Hayasaka {\em et al.}, ``{\em {Search for Lepton Flavor Violating Tau Decays into Three Leptons with 719 Million Produced Tau+Tau- Pairs}},'' \href{http://dx.doi.org/10.1016/j.physletb.2010.03.037}{Phys. Lett. B {\normalfont \bfseries 687} (2010)  139--143}, \href{http://arxiv.org/abs/1001.3221}{{\normalfont \ttfamily arXiv:1001.3221}}.

\end{thebibliography}\endgroup
\end{document}